\documentclass{ws-ijmpd}
\usepackage[nospace,sort]{overcite}

\def\lapp{\ifmmode\stackrel{<}{_{\sim}}\else$\stackrel{<}{_{\sim}}$\fi}
\def\gapp{\ifmmode\stackrel{>}{_{\sim}}\else$\stackrel{<}{_{\sim}}$\fi}
\def\degr{\ifmmode ^\circ \else$^\circ$\fi}
\def\fdg{\hbox{$.\!\!^\circ$}}

\begin{document}

\title{PULSARS AND GRAVITY}

\author{R. N. Manchester}

\address{CSIRO Astronomy and Space Science\\ 
Epping NSW 1710, Australia \\
dick.manchester@csiro.au }

\maketitle

\begin{abstract}
  Pulsars are wonderful gravitational probes. Their tiny size and
  stellar mass give their rotation periods a stability comparable to
  that of atomic frequency standards. This is especially true of the
  rapidly rotating ``millisecond pulsars'' (MSPs). Many of these
  rapidly rotating pulsars are in orbit with another star, allowing
  pulsar timing to probe relativistic perturbations to the orbital
  motion. Pulsars have provided the most stringent tests of theories
  of relativistic gravitation, especially in the strong-field regime,
  and have shown that Einstein's general theory of relativity is an
  accurate description of the observed motions. Many other
  gravitational theories are effectively ruled out or at least
  severely constrained by these results. MSPs can also be used to form
  a ``Pulsar Timing Array'' (PTA). PTAs are Galactic-scale
  interferometers that have the potential to directly detect nanohertz
  gravitational waves from astrophysical sources. Orbiting
  super-massive black holes in the cores of distant galaxies are the
  sources most likely to be detectable. Although no evidence for
  gravitational waves has yet been found in PTA data sets, the latest
  limits are seriously constraining current ideas on galaxy and
  black-hole evolution in the early Universe.
\end{abstract}


\section{Introduction}
Pulsars are rotating neutron stars that emit beams of radiation which
sweep across the sky as the star rotates. A beam sweeping across the
Earth may be detected as a pulse that repeats with a periodicity equal
to the rotation period of the star. Because of the large mass of
neutron stars, $\sim 1.4$~M$_\odot$, and their tiny size, radii $\sim
15$~km, (see Ref.~\refcite{lat12} for a review of neutron-star properties)
the rotation period of neutron stars is incredibly stable, with a
stability comparable to that of the best atomic clocks on Earth. This
great period stability, combined with the fact that pulsars are often
in a binary orbit with another star, makes them wonderful probes of
relativistic gravity. Tiny perturbations to their period resulting
from, for example, relativistic effects in a binary orbit, may be
detected and compared with the predictions of a gravitational
theory. Pulsars may also be used as detectors for gravitational waves
passing through the Galaxy. To separate the effects of gravitational
waves from other perturbations, signals from pulsars in different
directions on the sky must be compared -- exactly analogous to the way
laser-interferometer gravitational-wave detectors compare laser phases
in perpendicular arms.

More than 2400 pulsars are now known.\footnote{See the ATNF Pulsar
  Catalogue: http://www.atnf.csiro.au/research/pulsar/psrcat.} The
vast majority of them reside in our Galaxy, typically at distances of
a few thousand light-years from the Sun. Their pulse periods, $P$,
range between 1.4 milliseconds and 12 seconds and fall into two main
groups. The so-called ``normal'' pulsars have periods longer than
about 30 milliseconds and the ``millisecond'' pulsars (MSPs) have
shorter periods. MSPs comprise about 15\% of the known pulsar
population. 

\begin{figure}[hb]
\centerline{\includegraphics[width=90mm]{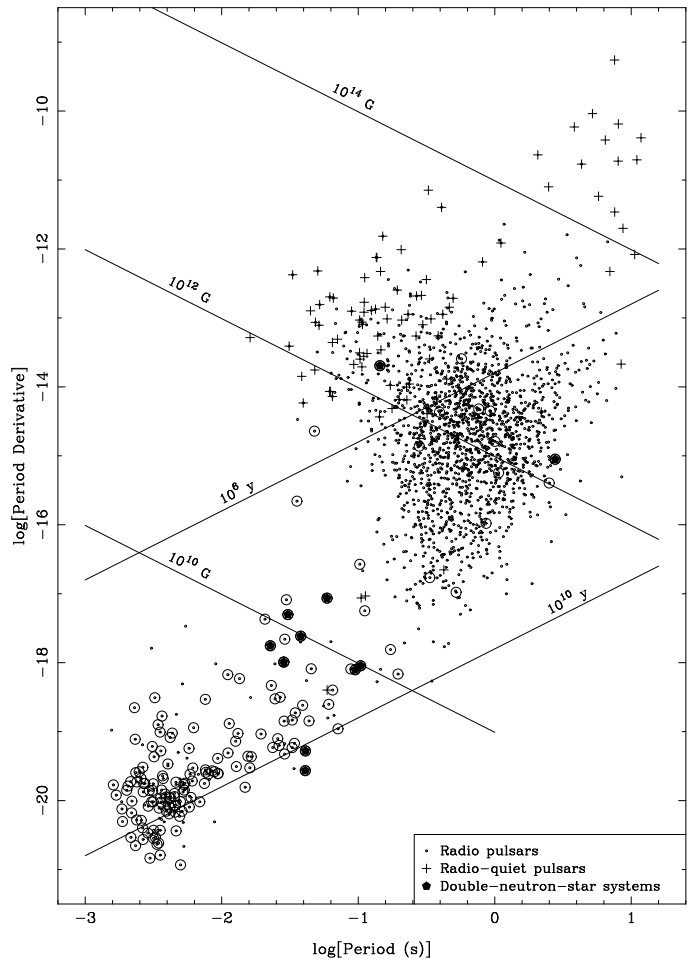}}
\caption{Distribution of pulsars on the $P$ -- $\dot P$ plane with
  radio-loud and radio-quiet pulsars indicated. Binary pulsars are
  indicated by a circle around the point and, for those with a
  neutron-star companion, the circle is filled in. Lines of constant
  characteristic age ($\tau_c$) and surface-dipole magnetic field
  ($B_s$) are shown. }
\label{fg:ppdot}
\end{figure}

Although pulsar periods are very stable, they are not
constant. In their own reference frame, all pulsars are slowing down,
albeit very slowly. Pulsars are powered by their rotational
energy. They have extremely strong magnetic fields and, as they spin,
they emit streams of relativistic particles and so-called ``dipole
radiation'', electromagnetic waves with period equal to the rotation
period of the star. These carry energy away from the star resulting in
a steady increase in the pulse period. Typical rates of period
increase, $\dot P$, are a part in $10^{15}$ for normal pulsars and
much less for MSPs. Assuming that the surrounding magnetic field is
predominantly dipole, the characteristic age $\tau_c$ of a pulsar is
given by
\begin{equation}
\tau_c = P/(2\dot P)
\end{equation}
and their surface dipole magnetic field strength (in
gauss\footnote{1~gauss (G) = $10^{-4}$~tesla}) is 
\begin{equation}
B_s \approx 3.2\times 10^{19} (P\dot P)^{1/2} \; {\rm G.}
\end{equation}
Figure~\ref{fg:ppdot} shows the distribution of pulsars
on the $P$ -- $\dot P$ plane, with several different types of pulsars
indicated. For most normal pulsars, $\tau_c$ is between $10^3$ and $10^7$ years
and $B_s$ is between $10^{11}$ and $10^{13}$~G. For MSPs, the
corresponding ranges are $10^9$ to $10^{11}$~yrs and $10^8$ to
$10^{10}$~G. 

Normal pulsars and MSPs have quite different evolutionary
histories. Most if not all normal pulsars are formed in supernova
explosions at the death of a massive star. They age with relatively
constant $B_s$ until the pulse emission mechanism begins to fail when
$\tau_c$ reaches about $10^6$~yrs. Many young pulsars ($\tau_c \lapp
10^4$~yrs) are located within supernova remnants, with the most famous
example being the Crab pulsar, PSR B0531+21, located near the centre
of the Crab Nebula. Most young pulsars lie relatively close to the
Galactic Plane, consistent with the idea that they are formed from
massive stars. MSPs on the other hand are much more widely distributed
in the Galactic halo. They are believed to originate from old, slowly
rotating and probably dead neutron stars that accrete matter and
angular momentum from an evolving binary companion. This ``recycling''
process increases their spin rate so that they have periods in the
millisecond range and re-energises the beamed emission (see, e.g.,
Ref.~\refcite{bv91}).

Figure~\ref{fg:ppdot} shows that the majority of MSPs are members of a
binary system, consistent with this formation scenario. The accretion
also suppresses their apparent magnetic field, so that MSPs have $\dot
P$ values about five orders of magnitude less than normal
pulsars. Since the level of intrinsic period irregularities is related
to $\dot P$,\cite{sc10} it is this low $B_s$ that makes MSPs extremely
stable clocks and suitable tools for the study of relativistic
gravitation. Figure~\ref{fg:ppdot} also indicates the class of
radio-quiet pulsars. Essentially all of the 72 known radio-quiet
pulsars are young and solitary. They include the so-called
``magnetars'' which lie in the upper right side of the $P$ -- $\dot P$
diagram where dipole magnetic fields are strongest.

Most double-neutron-star systems lie in the zone between the MSPs and
the normal pulsars. These systems are believed to have been partially
recycled prior to the formation of the second-born neutron star. In
almost every case, the observed pulsar is the recycled one since it
has a much longer active lifetime than the newly formed young
pulsar. A famous exception to this is the Double Pulsar (PSR
J0737$-$3039A/B) in which the second-born star (B) is (or was) still
visible \cite{lbk+04}. On Figure~\ref{fg:ppdot} the B pulsar is the
solitary double-neutron-star system on the right side of the plot. The
double-neutron-star system identified near the middle of the plot
contains the relatively young pulsar, PSR J1906+0746. There is some
doubt about the nature of the companion in this system -- it could be
a heavy white dwarf \cite{vks+15}. These double-neutron-star systems
and their use as probes of relativistic gravity are discussed in some
detail in Section~\ref{sec:dns} below.

\subsection{Pulsar Timing}\label{sec:timing}
The most important contributions of pulsars to investigations of
gravitational theories and gravitational waves rely on precision
pulsar timing observations. These allow both the relativistic
perturbations to binary orbits to be studied in detail and the potential
detection of the tiny period fluctuations generated by gravitational
waves passing through our Galaxy. Because of the importance of pulsar
timing to these studies, we give here a brief review of its basic
principles. 

The basic observable in pulsar timing is the time of arrival (ToA) of
a pulse at an observatory. In fact, because of signal/noise
limitations and the intrinsic fluctuation of individual pulse shapes,
ToA measurements are based on mean pulse profiles formed by
synchronously averaging the data, typically for times between several
minutes and an hour. The time at which a fiducial pulse phase (usually
near the pulse peak) arrives at the telescope is determined by
cross-correlating the observed mean pulse profile with a standard pulse
template. A series of these ToAs is measured over many days, months,
years and even decades for the pulsar of interest.

These observatory ToAs are affected by the rotational and
orbital motion of the Earth (and for satellite observatories, the
orbital motion of the satellite). To remove these effects, the
observed ToAs are referred to the solar-system barycentre (centre of
mass) which is assumed to be inertial (unaccelerated) with respect to the
distant Universe.\footnote{This neglects any acceleration of the
  solar-system barycentre resulting from, for example, Galactic
  rotation. For some precision timing experiments, such effects are
  taken into account at a later stage of the analysis.} This
correction makes use of a solar-system ephemeris giving predictions of
the position of the centre of the Earth with respect to the
solar-system barycentre. Such ephemerides, for example, the Jet
Propulsion Laboratory ephemeris DE~421,\cite{fwb09} are generated by fitting
relativistic models to planetary and spacecraft data. The correction
also takes into account the relativistic variations in terrestrial time
resulting from the Earth's motion. 

The resulting barycentric ToAs are then compared with predicted pulse
ToAs based on a model for the pulsar. The pulsar model can have 20 or more
parameters; generally included are the pulse frequency ($\nu = 1/P$),
frequency time derivative ($\dot\nu$), the pulsar position and the five
Keplerian binary parameters if the pulsar is a member of a binary
system. If the data are recorded at different radio frequencies, it is
also necessary to include the dispersion measure (DM) which quantifies
the frequency-dependent delay suffered by the pulses as they propagate
through the interstellar medium. 

\begin{figure}[ht]
\centerline{\includegraphics[width=85mm]{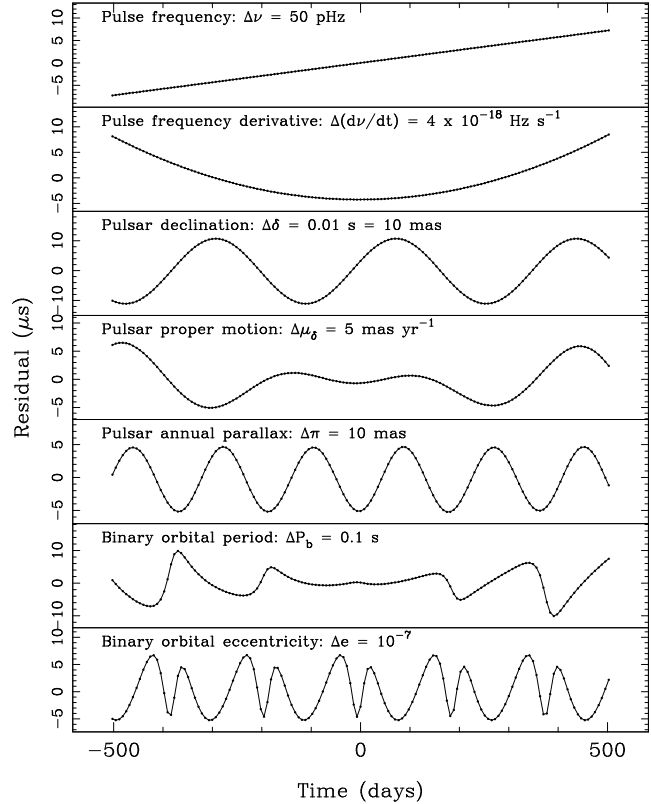}}
\caption{Variations in timing residuals for offsets in several
  parameters. The model pulsar has a pulse frequency of 300 Hz (3.3~ms
period) and is in an eccentric ($e = 0.4$) binary orbit of period 190
days. The reference epochs for period, position and binary phase are at
the middle of the plotted range. }
\label{fg:model_res}
\end{figure}

The differences between the observed and predicted ToAs are known as
``timing residuals''. Errors in any of the model parameters result in
systematic variations in the timing residuals as a function of
time. For example, if the model pulse frequency is too small, the
residuals will grow linearly as illustrated in
Figure~\ref{fg:model_res}. The required correction to the pulse
frequency is given by the slope of this variation. An error in the pulsar
position results in an annual sine curve which arises from
the barycentric correction. The phase and amplitude of this curve
give the corrections to the two position coordinates. Similarly, a pulsar
proper motion results in a linearly growing sine curve (away from the
reference epoch). For a sufficiently close pulsar, the curvature of the
wavefront results in a biannual term in the residuals, and offsets in
binary parameters result in terms which vary with the orbital
periodicity. 

Since offsets in different parameters in general result in different
systematic residual variations, it is possible to do a least-squares
fit to the observed residuals for the correction to any desired
parameter.\footnote{The data span must be sufficiently long to avoid
  excessive covariance between the variations for different fitted
  parameters.} The precision of the fitted parameters is often
amazingly high. For example, the relative precision of the pulse
frequency determination is $\sim \delta t /T$, where $\delta t$ is the
typical uncertainty in the ToAs and $T$ is the data span. Since, for
MSPs at least, $\delta t$ is often $<1$~$\mu$s and $T$ is often many
years, the relative precision of the measured $\nu$ can easily exceed
1:10$^{14}$. Similarly pulsar positions can be measured to
micro-arcseconds and binary eccentricities measured to $1:10^8$. These
very high precisions often allow higher-order terms, for example,
resulting from relativistic perturbations to the binary orbit, to be
measured.

\section{Tests of Relativistic Gravity}\label{sec:rel_grav}

\subsection{Tests of General Relativity with Double-Neutron-Star
  Systems}\label{sec:dns}
\subsubsection{The Hulse-Taylor binary, PSR B1913+16}\label{sec:B1913}
\begin{figure}[ht]
\centerline{\includegraphics[width=85mm]{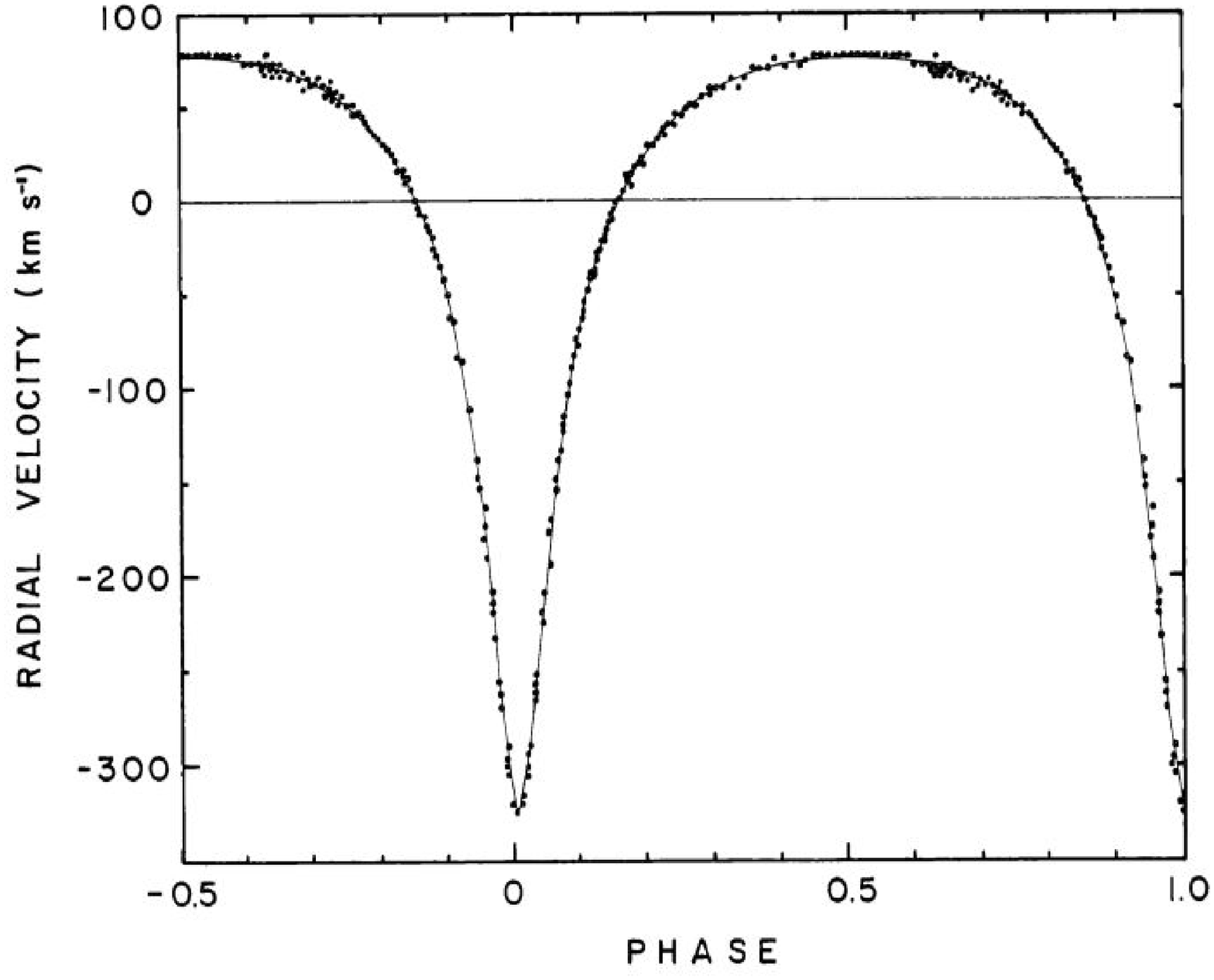}}
\caption{Velocity curve for the Hulse-Taylor binary pulsar,
  PSR~B1913+16. (Ref.~\protect\refcite{ht75a})}
\label{fg:B1913_vel}
\end{figure}

The discovery at Arecibo Observatory in 1974 of the first-known binary pulsar,
PSR B1913+16, by Hulse and Taylor\cite{ht75a} was remarkable in a number of
respects. Firstly, it showed that pulsars with short pulse periods but
large characteristic ages (59~ms and $10^8$~yr, respectively, for PSR
B1913+16) could exist. The period of PSR B1913+16 was second only to
the Crab pulsar, but its age was enormously greater than that of
other short-period pulsars known at the time. Secondly, it was in a
binary orbit with a relatively massive star, very likely another neutron
star, showing that an evolutionary pathway to such systems
existed. Thirdly, its orbital period was extraordinarily short, only
7.75~h, its eccentricity large, $\sim 0.617$, and, as shown in
Figure~\ref{fg:B1913_vel}, its maximum orbital velocity very high,
$\sim 300$~km~s$^{-1}$ or 0.1\% of the velocity of light. As was
immediately recognised by Hulse and Taylor, these properties opened up
the possibility that relativistic perturbations to the orbit were
potentially measurable. Lowest-order relativistic effects go as
$(v/c)^2$, and so the variations are of order 1:10$^6$, enormous
by the standards of pulsar measurements. 

Relativistic effects in binary motion can be expressed in terms of
``post-Keplerian'' parameters that describe departures from Keplerian
motion (see, e.g., Ref.~\refcite{sta03}). The first such parameter to
be observed was periastron precession \cite{thf+76}. In Einstein's
general theory of relativity (GR) the rate of periastron precession
(averaged over the binary orbit) is given by:
\begin{equation}\label{eq:pk-omdot}
\dot\omega = 3(P_b/2\pi)^{-5/3}(T_\odot M)^{2/3}(1-e^2)^{-1}
\end{equation}
where $\omega$ is the longitude of periastron measured from the
ascending node, $P_b$ is the orbital period, $T_\odot \equiv G
M_\odot/c^3 = 4.925490947 \mu$s, $G$ is the Gravitational Constant,
$M_\odot$ is the mass of the Sun, $e$ is the orbital eccentricity and
$M = m_1 + m_2$, the sum of the pulsar mass $m_1$ and the companion
mass $m_2$ in solar units. It is worth noting that this is the same
effect as the excess perihelion advance of Mercury that was used by
Einstein\cite{ein15} as an observational verification of GR. The
relativistic effect for Mercury is just 43 arcsec per century,
minuscule compared to the $4\fdg22$ per year predicted and observed
for PSR B1913+16.

The next most significant parameter, normally labelled $\gamma$, describes the
combination of gravitational redshift and 2nd-order (or transverse)
Doppler shift, both of which have the same dependence on orbital phase. In
GR
\begin{equation}\label{eq:pk-gamma}
\gamma = e(P_b/2\pi)^{1/3}T_\odot^{2/3}\;M^{-4/3}\;m_2(m_1+2m_2).
\end{equation}
Since the Keplerian parameters are very well determined, measurement
of $\dot\omega$ and $\gamma$ gives two equations in two unknowns,
$m_1$ and $m_2$, and so the two stellar masses can be determined. For
PSR B1913+16, these are both close to 1.4~$M_\odot$, confirming the
double-neutron-star nature of the system. An important consequence of
this was that the two stars could safely be treated as point masses in GR,
thereby allowing precise tests of the theory.

\begin{figure}[ht]
\centerline{\includegraphics[width=85mm]{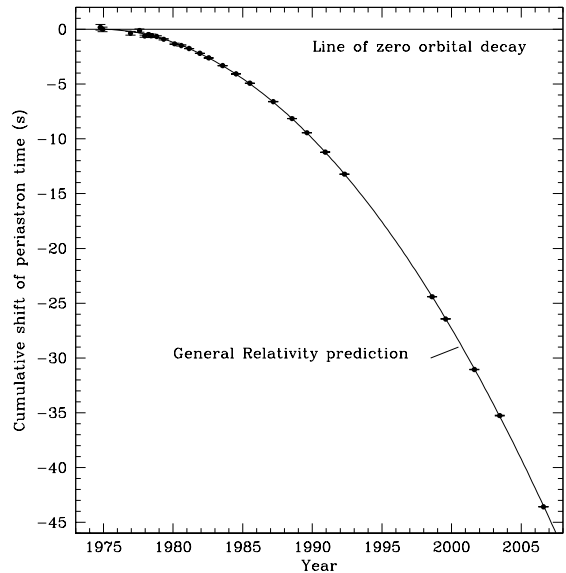}}
\caption{Comparison of the observed and predicted orbital period decay
  in PSR B1913+16. The orbital decay is quantified by the shift in the
  time of periastron passage with respect to a non-decaying orbit. The
  parabolic curve is the predicted decay from
  GR. (Ref.~\protect\refcite{wnt10})}
\label{fg:B1913_pbdot}
\end{figure}

Given the Keplerian parameters and the two masses, the next
post-Keplerian parameter, orbital decay due to the emission of
gravitational waves from the system, given in GR by
\begin{equation}\label{eq:pk-pbdot}
\dot P_b = -\frac{192\pi}{5} \left( \frac{P_b}{2\pi} \right)^{-5/3}
  \left( 1 + \frac{73}{24} e^2 + \frac{37}{96} e^4 \right) (1-e^2)^{-7/2}\;
  T_\odot^{5/3}\; m_1 m_2 M^{-1/3},
\end{equation}
could be predicted and compared with observation. This therefore
constituted a unique test of GR in strong-field gravity. After just
four years, the $\dot P_b$ term was well measured and
shown to be consistent with the GR prediction \cite{tfm79}. The orbital decay,
including more recent results, is illustrated in
Figure~\ref{fg:B1913_pbdot}; the ratio of observed to predicted orbit
decay is $0.997\pm 0.002$ \cite{wnt10}. It should be mentioned that
this near-perfect agreement relies on correcting the observed $\dot
P_b$ for the differential acceleration of the solar system and the
pulsar system in the gravitational field of the Galaxy. This
correction is $-0.027\pm 0.005\times 10^{-12}$ or about 1\% of the
observed value, and the uncertainty in the final result is dominated
by the uncertainty in this correction. Unfortunately, it is unlikely
that this uncertainty will be significantly reduced in the near future
since it mainly depends on the poorly known distance to the binary
system. 

Two more post-Keplerian parameters, denoted by $r$ and $s$,
relate to the Shapiro delay suffered by the pulsar signal while
passing through the curved spacetime surrounding the companion
star.\cite{sha64} The relations for them in GR are as follows:
\begin{equation}\label{eq:pk-r}
r = T_\odot m_2 
\end{equation}
\begin{equation}\label{eq:pk-s}
s \equiv \sin i = (a_1/c) \sin i\; \left(\frac{P_b}{2\pi}\right)^{-2/3} 
    T_\odot^{-1/3} M^{2/3} m_2^{-1} 
\end{equation}
where $(a_1/c) \sin i$ is the projected semi-major axis of the pulsar
orbit (in time units), a Keplerian parameter. The Shapiro delay is
generally only observable when the orbital inclination is relatively
close to 90\degr, that is, the orbit is seen close to edge-on. For the
Hulse-Taylor binary pulsar the orbital inclination is about 47\degr
and the Shapiro delay is small and covariant with Keplerian
parameters. However it has been observed in a number of other
neutron-star binary systems as will be discussed below.

Another relativistic effect, geodetic precession, has observable
consequences for PSR B1913+16 and several other binary pulsars. A
neutron star formed in a supernova explosion receives a ``kick''
during or shortly after the explosion which typically gives the star a
velocity of several hundred km~s$^{-1}$ \cite{hllk05}. If the pulsar
is a member of a binary system which is not disrupted by the kick, the
orbital axis is changed so that pulsar spin axis, which was most
likely aligned with the orbital axis prior to the explosion, is no
longer aligned. Since kick velocities are often comparable to or even
larger than the orbital velocities, the misalignment angle can be
large. Precession of the spin axis will therefore alter the aspect of
the radio beam seen by an observer and may even move the beam out of
the line of sight.

\begin{figure}[ht]
\centerline{\includegraphics[width=85mm]{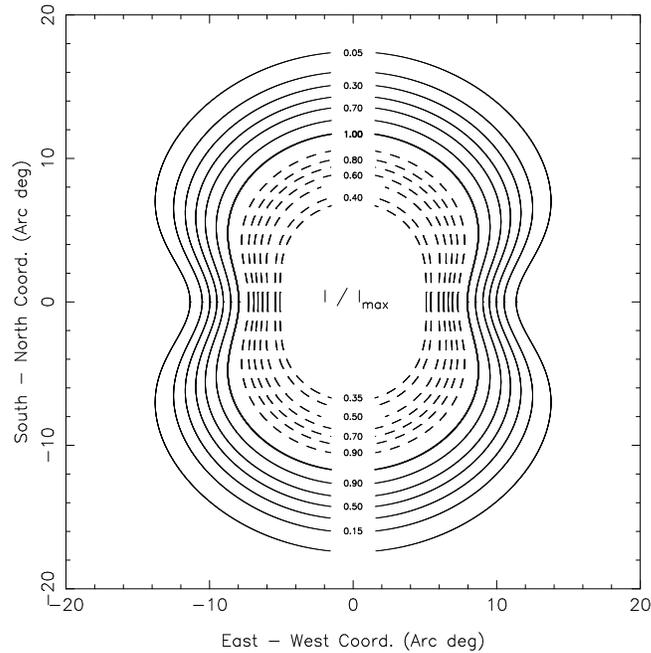}}
\caption{Contours of the symmetric part of the radio emission beam for PSR
  B1913+16, obtained by fitting to the variation in pulse width over the
  interval 1981 to 2001. (Ref.~\protect\refcite{wt02})}
\label{fg:B1913_beam}
\end{figure}

In GR, the precessional angular frequency is given by
\begin{equation}\label{eq:geo_prec}
\Omega_p = \frac{1}{2}\left(\frac{P_b}{2\pi}\right)^{-5/3} 
    T_\odot^{2/3} \frac{m_2(4m_1 + 3m_2)}{(1-e^2)M^{4/3}}.
\end{equation}
For PSR B1913+16, the corresponding precessional period is about 297
years, so the aspect changes over observational data spans are
small. Never-the-less changes in the relative amplitude of the two
peaks in the PSR B1913+16 profile were reported by Weisberg et
al.\cite{wrt89} and attributed to the effects of geodetic precession
with a ``patchy'' beam. For a basically conal beam geometry, the
separation of the two profile components would be expected to change
and evidence for this was first found by Kramer\cite{kra98} leading to
an estimate of the misalignment angle of about 22\degr. A data set
extending to 2001 was analysed by Weisberg and Taylor\cite{wt02},
obtaining results similar to those of Kramer\cite{kra98}.  Their
best-fitting model gives the ``peanut'' shaped beam shown in
Figure~\ref{fg:B1913_beam}. Clifton and Weisberg\cite{cw08} have shown
that a set of circular nested emission cones can also give apparent
pulse-width variations similar to those observed. Over the 20-year
interval covered by the data set, the ``impact parameter'' (minimum
angle between the beam symmetry axis and the observer's line of sight)
has changed by a rather small amount, about 3\degr, from $-3\fdg5$ to
$-6\fdg5$. Extrapolation of this model suggests that the pulsar will
become unobservable in about 2025. While this result is compatible
with relativistic precession, it is not possible to derive an
independent measure of the precessional rate from these data.

\subsubsection{PSR B1534+12}\label{sec:B1534}
\begin{figure}[ht]
\centerline{\includegraphics[width=85mm]{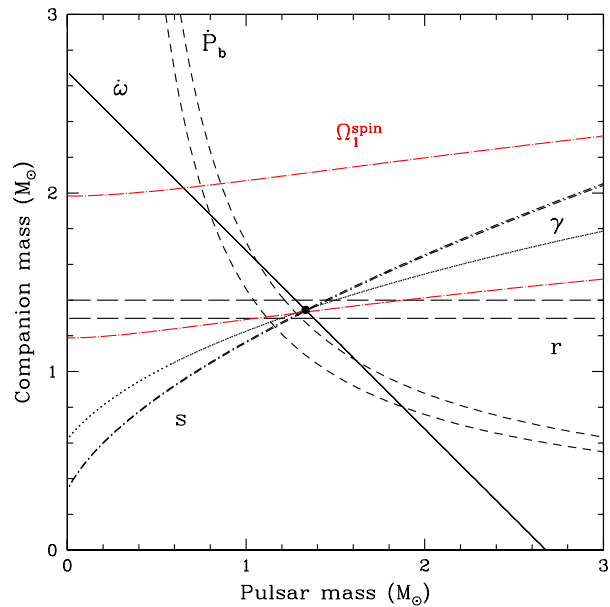}}
\caption{Plot of companion mass ($m_2$) versus pulsar mass ($m_1$) for
  PSR B1534+12. Constraints derived from the measured post-Keplerian
  parameters assuming GR are shown pairs of lines with separation
  indicating the uncertainty range. For $\dot\omega$ and $\gamma$, the
  uncertainty ranges are too small to see on the
  plot. In addition, a constraint from the measured spin precession
  rate is shown.  (Ref.~\protect\refcite{fst14})}
\label{fg:B1534_m1m2}
\end{figure}

PSR B1534+12, discovered by Wolszczan in 1990,\cite{wol91a} is a
binary system with parameters quite similar to those of B1913+16,
notably a short orbital period ($\sim 10.1$~hours), relatively high
eccentricity ($\sim 0.27$) and a compact orbit about 60\% larger than
that of PSR B1913+16. Analysis of less than a year's data already
allowed measurement of two post-Keplerian parameters, $\dot\omega$ and
$\gamma$, thereby determining the masses of the two stars and
confirming that they are both neutron stars. These results also showed
that the orbit was more edge-on than that of PSR B1913+16, with an
implied inclination angle of about 77\degr. Analysis of timing data
extending over 22 years by Fonseca et al.\cite{fst14} built on earlier
results by Stairs et al.\cite{sttw02}, with significant detections of
$r$ and $s$, the Shapiro-delay parameters, and the orbital decay,
$\dot P_b$, giving five post-Keplerian parameters and three
independent tests of GR. A fourth test of GR, albeit less precise, was
provided by an analysis of the evolution of the profile shape and
polarisation, yielding a rate for the geodetic precession of the
pulsar spin axis $\Omega_p = 0\fdg59^{+0.12}_{0.08}$~yr$^{-1}$ which is
consistent with the value predicted by GR. Figure~\ref{fg:B1534_m1m2}
shows the so-called ``mass-mass'' diagram for PSR B1534+12,
illustrating these constraints.

If GR gives a correct description of the post-Keplerian parameters,
all of these constraints should be consistent with an allowed range of
$m_1$ and $m_2$. For PSR B1534+12, the masses are most accurately
constrained by $\dot\omega$ and $\gamma$ but the predicted constraint
on $\dot P_b$ appears to be inconsistent. As for PSR B1913+16, the
measured value of $\dot P_b$ must be corrected for kinematic effects
resulting from the differential acceleration of the binary and solar
system in the Galaxy. PSR B1534+12 is much closer to the Sun that PSR
B1913+16 and the so-called ``Shklovskii'' effect,\cite{shk70} an
apparent radial acceleration resulting from transverse motion of the
binary system ($\dot P/P = \mu^2d/c$, where $\mu$ is the pulsar proper
motion and $d$ is the pulsar distance), is also important. The
distance estimate is based on the pulsar DM and a model for the free
electron density in the Galaxy and is quite uncertain. Stairs and her
colleagues \cite{sttw02,fst14} inverted the problem, assuming that the
GR prediction for $\dot P_b$ is correct, thereby deriving an improved value
for the pulsar distance, a technique first suggested by Bell and
Bailes \cite{bb96}.

\subsubsection{The Double Pulsar, PSR J0737$-$3039A/B}\label{sec:J0737}
The discovery of the Double Pulsar system \cite{bdp+03,lbk+04}
heralded a remarkable era for investigation of relativistic effects in
double-neutron-star systems. In this system, the A pulsar was formed
first and subsequently spun up to approximately its current period of
23~ms by accretion from its evolving binary companion. The companion
then imploded to form the B pulsar which has since spun down to its
current period of about 2.8~s. The orbital period is only 2.5 hours,
less than a third of that for PSR B1913+16, and the projected
semi-major axis $a_1 \sin i$ is about 60\% that of PSR B1913+16. These
parameters imply relativistic effects much larger than those seen for
PSR B1913+16; for example, the predicted relativistic periastron
advance is 16\fdg9~yr$^{-1}$, more than four times the value for PSR
B1913+16. Added to that, the orbit is seen within a degree or so of
edge-on, not only allowing detailed measurement of the Shapiro delay,
but also resulting in eclipses of the A pulsar emission by the
magnetosphere of the B pulsar. Finally, the still-unique detection of
the second neutron star as a pulsar allows a direct measurement of the
mass ratio of the two stars from the ratio of the two non-relativistic
Roemer delays ($a_1 \sin i$).\footnote{The B pulsar became
  undetectable as a radio pulsar in 2008, most probably because the
  beam precessed out of our line of sight.\cite{pmk+10} Because of
  uncertainty about the beam shape, the date of its return to
  visibility is very uncertain, but it should be before 2035.}

\begin{figure}[ht]
\centerline{\includegraphics[angle=270,width=85mm]{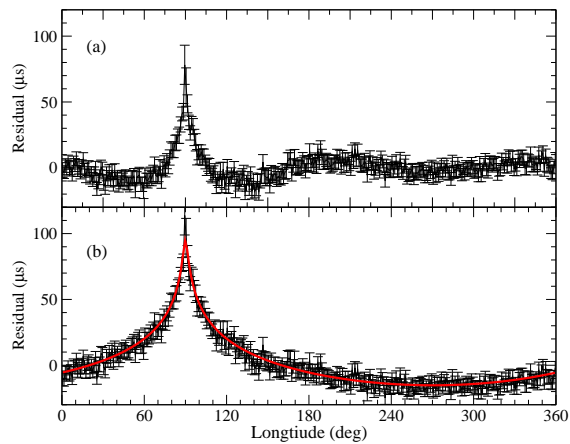}}
\caption{Observed Shapiro delay as a function of orbital phase for PSR
  J0737$-$3039A. Upper panel: timing residuals after fitting for all
  parameters except the Shapiro-delay terms, $r$ and $s$, with these
  set to zero. Lower panel: the full Shapiro delay obtained by taking
  the best-fit values from a full solution, but setting $r$ and $s$ to
  zero. The red line is the prediction based on GR.
  (Ref.~\protect\refcite{ksm+06})}
\label{fg:shapiro}
\end{figure}

\begin{figure}[ht]
\centerline{\includegraphics[angle=270,width=85mm]{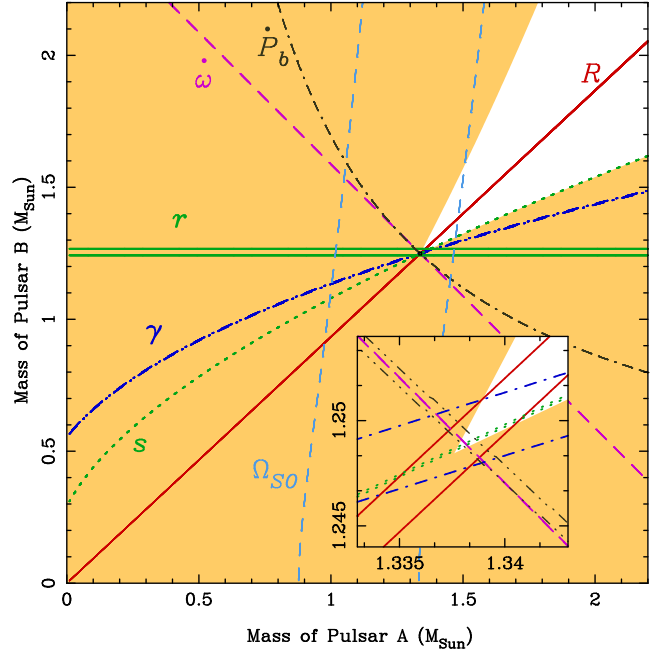}}
\caption{Plot of companion mass ($m_2$) versus pulsar mass ($m_1$) for
  PSR J0737$-$3039 with observed constraints interpreted in the
  framework of GR. The inset shows the central region at an expanded
  scale, illustrating that GR is consistent with all
  constraints. (M. Kramer, private communication) }
\label{fg:J0737_m1m2}
\end{figure}

Four post-Keplerian parameters ($\dot\omega$, $\gamma$, $r$ and $s$)
were detected in just seven months of timing data from the Parkes 64-m
and Lovell 76-m Telescopes \cite{lbk+04}. Further observations,
including data from the Green Bank 100-m Telescope, with a 2.5-year
data span give the currently most stringent test of GR in strong-field
conditions \cite{ksm+06,kw09}. Figure~\ref{fg:J0737_m1m2} shows the
mass-mass diagram based on these results together with the measurement
of geodetic spin precession for the B pulsar described below. A total
of six post-Keplerian parameters together with the mass ratio $R$
gives five independent tests of GR. As well as the three
post-Keplerian parameters, $\dot\omega$, $\gamma$ and $\dot P_b$
observed for PSRs~B1913+16 and PSR B1534+12
(Figure~\ref{fg:B1534_m1m2}), for the Double Pulsar we have the mass
ratio $R$, the Shapiro delay parameters $r$ and $s$ and a measurement
of the rate of geodetic precession $\Omega_p$. The
observed Shapiro delay (Figure~\ref{fg:shapiro}) shows that the
J0737$-$3039A/B orbit inclination angle is $88\fdg7 \pm 0\fdg7$.

As mentioned above, this nearly edge-on view of the orbit results in
eclipses of of the A-pulsar emission by the magnetosphere of the B
pulsar. These eclipses last only about 30~s, showing that the B-pulsar
magnetosphere is highly modified by the relativistic wind from pulsar
A \cite{lbk+04}. Remarkably, observations with high time resolution
made with the Green Bank Telescope showed that the eclipse is
modulated at the spin period of pulsar B \cite{mll+04}. Modelling of
the detailed eclipse profile by absorption in the doughnut-shaped
closed-field-line region of the magnetosphere by Lyutikov and
Thompson\cite{lt05} allowed determination of the geometry of the
binary system including the offset between the B-pulsar spin axis and
the orbital angular momentum axis, the so-called ``misalignment
angle'' which they estimate to be about 60\degr. Even more remarkably,
detailed measurements of the eclipse profile over about four years
enabled Breton et al.\cite{bkk+08} to directly estimate the rate of geodetic
precession as $4\fdg77 \pm 0\fdg66$ yr$^{-1}$, consistent with the
predicted precessional period of 70.95 years based on GR
(Equation~\ref{eq:geo_prec}). It is notable that no secular profile
evolution has been observed for PSR J0737$-$3039A, implying that,
unlike for the B pulsar, the misalignment angle for pulsar A is very
small \cite{fsk+13}.

The most precisely measured post-Keplerian parameter is
$\dot\omega$. This constraint is nearly orthogonal to that from the
mass ratio $R$ (Figure~\ref{fg:J0737_m1m2}) giving values for the two
neutron-star masses of $m_1 = 1.3381\pm 0.0007$~M$_\odot$ and $m_2 =
1.2849\pm 0.0007$~M$_\odot$ \cite{ksm+06}. Together with the accurately measured
Keplerian parameters, these two masses can be used to predict the
remaining five post-Keplerian parameters using
Equations~\ref{eq:pk-gamma} -- \ref{eq:geo_prec}. As
Figure~\ref{fg:J0737_m1m2} shows, GR gives a self-consistent
description of the orbital motions, with all five independent
constraints consistent with the masses derived from $\dot\omega$ and
$R$.

The most stringent test comes from the derived value of $s$,
$0.99974\pm 0.00039$. The ratio of the observed and predicted values
of $s$ is $0.99987\pm 0.0005$, by far the most constraining test of GR
in the strong-field regime. Furthermore, this test is qualitatively
different to that for PSR B1913+16 in that it is based on a
non-radiative prediction. With just 2.5 years of data analysed, the
orbital decay term for the Double Pulsar is not measured as precisely
as that for PSR B1913+16. However, since the phase offset grows
quadratically (Figure~\ref{fg:B1913_pbdot}) and the effect of receiver
noise decreases as $T^{1/2}$ where $T$ is the data span (assuming
approximately uniform sampling), the precision of the $\dot P_b$
measurement should improve as $T^{5/2}$. Furthermore, J0737$-$3039A/B
is much closer to the Sun than PSR B1913+16, thus reducing the
magnitude and uncertainty of the correction due to differential
Galactic acceleration. Also, using very long-baseline interferometry
(VLBI), Deller et al.\cite{dbt09} have shown that the transverse space
velocity of the binary system is small, just
$24_{-6}^{+9}$~km~s$^{-1}$, and so the Shklovskii correction to the
observed $\dot P_b$ is also small and accurately known. For these
reasons, the orbit-decay test with the Double Pulsar will not be
limited by these corrections, at least for another decade or
so. Beyond that, improved models for the Galactic gravitational
potential can be expected from analysis of GAIA data \cite{rb13a} and
improved parallax and proper motion measurements may come from further
VLBI observations, both reducing the uncertainty in these kinematic
corrections.

Several other post-Keplerian parameters can in principle be observed
in binary systems \cite{dt92}, testing other aspects of
relativity. Kramer and Wex\cite{kw09} show that only one of these,
$\delta_\theta$, which quantifies relativistic deformations of the
binary orbit, is potentially measurable with the Double Pulsar
system. Even this will take more than a decade to reach an interesting
level of precision given reasonable expectations about future
observations.

A potentially exciting prospect is to use measurement of higher-order
terms for relativistic periastron precession to put constraints on the
moment of inertia of a neutron star \cite{ds88}. At the second
post-Newtonian level, the periastron precession has three components:
\begin{equation}\label{eq:omdot_2pn}
\dot\omega = \dot\omega_{1pN} + \dot\omega_{2pN} + \dot\omega_{SO}
\end{equation}
where $\dot\omega_{1pN}$ is given by Equation~\ref{eq:pk-omdot},
$\dot\omega_{2pN}$ is the 2nd post-Newtonian $(v/c)^4$ contribution
and $\dot\omega_{SO}$ gives the contribution from spin-orbit
coupling \cite{bo75b,ds88}. This latter term depends on the moment of
inertia and spin of the A pulsar (the spin angular momentum of the B pulsar is
negligible) and can in principle be measured if the two leading
components in Equation~\ref{eq:omdot_2pn} can be measured to
sufficient precision. Kramer and Wex\cite{kw09} show that, given
expected advances in precision timing and astrometry, a
significant result could be obtained in 20 years or so.

Spin-orbit coupling can also lead to a nonlinear variation in $\omega$
as a function of time and a time variation in the projected semi-major
axis of pulsar A, $a_1 \sin i$ \cite{kw09}. However, the nonlinear
terms depend on $\sin \theta$, where $\theta$ is the misalignment
angle. As mentioned above, it seems as though $\theta_A$ is very
small, so unfortunately these terms may be difficult to detect.

\subsubsection{Measured post-Keplerian Parameters}\label{sec:pk_params}
Table~\ref{tb:3pk} summarises the measured post-Keplerian parameters
for the eleven pulsar binary systems where three or more post-Keplerian
parameters have been measured. Since only two measurements are
required to determine the stellar masses, these systems provide the
opportunity for tests of theories of relativistic gravity. Of the eleven
systems, in six cases the companion star is believed to be a neutron
star, in a further four cases a white dwarf companion is more likely
and in one case the companion is probably a main sequence star. This
latter system, PSR J1903+0327, evidently has had an unusual
evolutionary history as a triple system in which one component was
ejected \cite{fbw+11}. It is currently not clear if there is a
significant contribution to the observed $\dot\omega$ from kinematic
effects and/or spin-orbit coupling, so the utility of this system for
tests of relativistic gravity is limited. There a further ten pulsar
systems in the literature where just two post-Keplerian parameters
have been measured, enabling estimates of the stellar masses, but no
tests of relativistic gravity.

Although an independent measurement of relativistic spin precession
has been possible for just two systems so far, as shown in
Table~\ref{tb:3pk}, secular changes in observed pulse profiles have
been observed for four other pulsars and attributed to geodetic
precession of the pulsar spin axis.
\begin{table}[ht]
\tbl{Binary pulsars with three or more significant measured post-Keplerian parameters}
{\begin{tabular}{lcccccc} \toprule
Pulsar/Parameter & J0437$-$4715 & J0737$-$3039A/B & J0751+1807 & J1141-6545 & B1534+12 & J1756$-$2251\\
\colrule
Peri. advance $\dot\omega$ (\degr yr$^{-1}$) & 0.016(8) & 16.8995(7) & -- & 5.3096(4) & 1.7557950(19) & 2.58240(4) \\
Time dilation $\gamma$ (ms) & -- & 0.3856(26) & -- & 0.773(11) & 2.0708(5) & 0.001148(9) \\
Orb.P deriv. $\dot P_b$ ($10^{-12}$) & 3.73(6)$^{\text b}$  & $-1.252(17)$  &
$-0.031(5)$  & $-0.403(25)$ & $-0.1366(5)$ & $-0.229(5)$ \\
$s \equiv \sin i$ & 0.6746(28)$^{\text b}$  & 0.99974(39) & 0.90(5) &--  & 0.9772(16) & 0.93(4) \\
Comp. mass $m_2$ ($M_\odot$) & 0.254(18) & 1.2489(7) & 0.191(15) & -- & 1.35(5) & 1.6(6) \\
Geod. prec. $\Omega_p$ (\degr yr$^{-1}$) & -- & 4.77(66)$^{\text c}$ & 
-- & Note d & 0.59(10)  & -- \\
Binary companion$^{\text a}$ & He WD & NS & He WD & CO WD & NS & NS \\
References & \refcite{vbv+08} & \refcite{ksm+06,bkk+08} & \refcite{nss+05,nsk08} & \refcite{bbv08} & \refcite{fst14} & \refcite{fsk+14} \\
\colrule
Pulsar/Parameter & J1807-2459B & J1903+0327 & J1906+0746 & B1913+16 & B2127+11C & \\
\colrule
Peri. advance $\dot\omega$ (\degr yr$^{-1}$) & 0.018339(4) &
0.0002400(2)$^{\text b}$ & 7.5841(5) & 4.226598(5) & 4.4644(1) & \\
Time dilation $\gamma$ (ms) & 26(14) & -- & 0.470(5) & 4.2992(8) & 4.78(4) & \\
Orb.P deriv. $\dot P_b$ ($10^{-12}$) & -- & -- & $-0.56(3) $ & $-2.423(1)^{\text b}$ & $-3.96(5)$  & \\
$s \equiv \sin i$ & 0.99715(20) & 0.9759(16)  & -- & -- & --  &  \\
Comp. mass $m_2$ ($M_\odot$) & 1.02(17) & 1.03(3) & -- & -- & -- & \\
Geod. prec. $\Omega_p$ (\degr yr$^{-1}$)  & -- &-- & Note d & Note d & Note d & \\
Binary companion$^{\text a}$ & CO WD(?) & MS & NS(?) & NS & NS & \\
References & \refcite{lfrj12} & \refcite{fbw+11} & \refcite{vks+15,dkc+13} & \refcite{wt02,wnt10} & \refcite{jcj+06} & \\
\botrule
\end{tabular}
}
\begin{tabnote}
  a: Binary companion types: CO WD: Carbon-Oxygen White Dwarf; He WD:
  Helium White Dwarf; MS: Main-sequence star; NS: Neutron star\\
  b: Dominated or significantly biased by kinematic effects \\
  c: For PSR J0737$-$3039B \\
  d: Effects of precession observed, but no independent determination
  of $\dot\Omega_p$
\end{tabnote}
\label{tb:3pk}
\end{table}

\subsection{Tests of Equivalence Principles and Alternative Theories of
  Gravitation}\label{sec:EP} 
Pulsars and especially binary pulsars have unique advantages in
testing theories of relativistic gravitation as a result of their
often rapid spin, short orbital periods and the ultra-high density of
the under-lying neutron stars. As we have shown above, GR has been
amazingly successful in describing all measurements to
date. Never-the-less, investigations of quantum gravity and cosmology
suggest that, in some regimes, extensions or modifications of GR may
be required. This strongly motivates a search for departures from GR
within existing experimental capabilities. 

Gravitational theories have Equivalence Principles at their heart. The
Weak Equivalence Principle (WEP) is basic to Newtonian gravity,
stating that acceleration in a gravitational field is independent of
mass or composition. The Strong Equivalence Principle (SEP) adds
Lorentz invariance (no preferred reference frame) and position
invariance (no preferred location) for both gravitational and
non-gravitational interactions. GR satisfies the SEP whereas other
theories may violate the SEP or even the WEP in one or more respects.

Comparison of different gravitational theories has been greatly
facilitated by the ``parametrized post-Newtonian'' (PPN) formalism
which describes observable or potentially observable phenomena in a
theory-independent way. This formalism was first developed by Will and
Nordtvedt \cite{wn72} for ``weak-field'' situations, that is, where
$\epsilon \sim GM/Rc^2 \ll 1$, where $G$ is the Newtonian
gravitational constant, $M$ and $R$ characterise the size and mass of
the object and $c$ is the velocity of light. Many gravitational tests
are performed within the solar system where $\epsilon \lapp 10^{-5}$,
firmly in the weak-field regime. However in the vicinity of a neutron
star $\epsilon \sim 0.2$ and so strong-field effects are potentially
important. A number of authors have considered the generalisation of
the PPN formalism to strong-field situations (see, e.g., Refs
\refcite{de92,wil93,dt92}) allowing investigation of these effects.

Some Lorentz-violating theories predict a dependence of the velocity
of light on photon energy or polarisation \cite{wil14}. Pulsar
observations can be used to limit these theories, but potentially
stronger limits come from observations of gamma-ray bursts, polarised
extra-galactic radio sources and the cosmic microwave background
\cite{ni10}.

A recent comprehensive review of observational limits on theories of
gravitation by C.~M. Will can be found in Ref.~\refcite{wil14}. Pulsar
tests of relativistic gravitation have been reviewed by
I.~H. Stairs\cite{sta03} and, more recently, by N.~Wex.\cite{wex14}
Further details on many of the topics discussed here may be found in
these reviews.

\subsubsection{Limits on PPN parameters}\label{sec:ppn_limits}

The standard PPN formalism has ten parameters: $\gamma_{\rm PPN}$,
$\beta_{\rm PPN}$, $\xi$, $\alpha_1$, $\alpha_2$, $\alpha_3$, $\zeta_1$,
$\zeta_2$, $\zeta_3$ and $\zeta_4$. In GR $\gamma_{\rm PPN}$,
describing space curvature per unit mass, and $\beta_{\rm PPN}$, describing
superposition of gravitational fields, are unity and all others are
zero. $\xi$ describes preferred location effects, the $\alpha_n$
preferred frame effects and the others describe violations of momentum
conservation. ($\alpha_3$ also may be non-zero in this case.) Pulsar
observations do not directly constrain $\gamma_{\rm PPN}$ or $\beta_{\rm PPN}$
but are important in constraining many of the remaining PPN parameters
and in fact currently place the strongest constraints on several
parameters.

Damour and Sch\"afer \cite{ds91} recognised that wide-orbit
low-eccentricity binary pulsars, which generally have a white-dwarf
companion, could provide a valuable test of the SEP through a
strong-field extension of the solar-system tests pioneered by
Nordtvedt\cite{nor68b}. A violation of the SEP would cause bodies with different
gravitational self-energy to fall at different rates in an external
gravitational field. In a pulsar -- white-dwarf binary system, this
results in a forced eccentricity in the direction of the gravitational
field, that of the Galaxy in this case. This eccentricity is given by
\begin{equation}\label{eq:ecc_f}
{\bf e}_{\rm F} = \frac{3}{2} \frac{\Delta {\bf g}_\perp}
     {\dot\omega a (2\pi/P_b)}
\end{equation}
where ${\bf g}_\perp$ is the projection of the Galactic gravitational
field on to the orbital plane, $\dot\omega$ is the relativistic
periastron advance, $a$ is the orbit semi-major axis and $P_b$ the
orbital period. The ratio of the gravitational mass $m_g$ and the
inertial mass $m$ (which is exactly one if the SEP is obeyed) for
body $i$ is
described by 
\begin{equation}\label{eq:Delta}
\left(\frac{m_g}{m}\right)_i = 1 + \Delta_i = 1 + \eta_N(E_g/mc^2)_i + 
\eta_N^\prime(E_g/mc^2)_i^2 + ...
\end{equation}
where $\eta_N$ is the Nordtvedt parameter, a function of several PPN
parameters (see, e.g., Ref.~\refcite{wil14}) and $\Delta =
\Delta_1 - \Delta_2$. Violations of the SEP will result in non-zero
$\Delta$.

Since Damour and Sch\"afer \cite{ds91} first proposed this method of
testing SEP violations, the number of suitable pulsar -- white-dwarf
binary systems has increased greatly. Gonzalez et al.\cite{gsf+11}
combined data for 27 systems to place a 95\% confidence upper limit on
$|\Delta|$ of $4.6\times 10^{-3}$.\footnote{This limit may be
  slightly under-estimated -- see Wex.\cite{wex14}} Since $E_g/mc^2
\sim 0.1$ for a neutron star, this limit is not as strong as the
weak-field limit on $\eta_N \lapp 3\times 10^{-4}$ from lunar-laser
ranging \cite{wtb12}. However, it does enter the strong-field regime
and test possible violations of the SEP that solar-system tests cannot
reach.

The recent discovery by Ransom et al.\cite{rsa+14} of a remarkable
triple system containing a pulsar, PSR J0337+1715, and two white-dwarf
stars in essentially coplanar orbits, one in a relatively tight
1.6-day orbit with the pulsar and the other in a much wider 327-day
orbit around the inner system, opens up the possibility of a much more
sensitive test of the SEP. Precise timing observations of the pulsar
have already shown that the motion of the inner system is strongly
affected by the gravitational field of the outer white dwarf. The
gravitational field of this star, which has a mass of about
$0.41$~M$_\odot$, at the inner system is at least six orders of
magnitude larger than the Galactic gravitational field at the position
of a typical pulsar and so the effect of SEP violations may be
expected to be correspondingly larger\cite{fkw12}. Observations over
several orbital periods of the outer star will almost certainly be
necessary to isolate any SEP-related effects from other orbit
perturbations.

The wide-orbit low-eccentricity binary pulsars can also be used test
for violations of local Lorentz invariance (LLI) of the gravitational
interaction and momentum conservation that are described by the
parameter $\alpha_3$ and its strong-field generalisation
$\hat\alpha_3$. Bell and Damour \cite{bd96} show that such violations
produce a forced eccentricity analogous to that produced by the
Nordtvedt effect given by
\begin{equation}
|{\bf e}_F| = \hat\alpha_3 \frac{c_p|{\bf w}|}{24\pi}\frac{P_b^2}{P}
    \frac{c^2}{G(m_1 + m_2)} \sin\beta
\end{equation}
where $c_p$ is a dimensionless ``compactness'' parameter, for
neutron stars about 0.2 \cite{de92}, and $\beta$ is the angle
between ${\bf w}$, the velocity of the system with respect to a
reference frame defined (for example) by the cosmic microwave
background, and the pulsar's spin axis. A similar analysis to that for
the generalised Nordtvedt effect resulted in a 95\% confidence
limit of $|\hat\alpha_3| < 4.0\times 10^{-20}$, some 13 orders of
magnitude lower than the best solar-system test \cite{sfl+05}. It is
worth noting that the observed small scatter in the period derivatives of
MSPs had already been used by Bell\cite{bel96} to limit $|\alpha_3|$ to
$< 5\times 10^{-16}$. 

\begin{figure}[ht]
\centerline{\includegraphics[width=85mm]{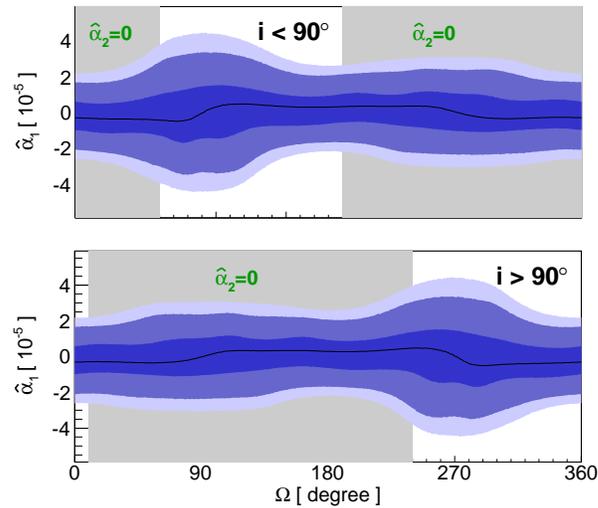}}
\caption{Constraints on the strong-field PPN parameter $\hat\alpha_1$
  from timing observations of the low-eccentricity binary pulsar PSR
  J1738+0333. The constraints are a function of the unknown
  orientation of the binary system on the sky (described by the
  longitude of the ascending node $\Omega$) and the unknown sign of
  $\sin i$. If the PPN parameter $\hat\alpha_2$ is assumed to be zero,
  then only certain ranges of $\Omega$ are permitted. The shading
  corresponds to 68\%, 95\% and 99.7\% confidence limits on
  $\hat\alpha_1$. (Ref.~\protect\refcite{sw12}) }
\label{fg:alpha1}
\end{figure}

Strong-field limits on the other two PPN parameters describing
preferred frame effects, specifically LLI, $\hat\alpha_1$ and
$\hat\alpha_2$, can also be obtained from low-eccentricity
binary pulsar observations \cite{de92}. Non-zero $\hat\alpha_1$
induces a forced eccentricity in the direction of motion of the
binary system velocity ${\bf w}$, analogous to Nordtvedt $\hat\alpha_3$ tests
discussed above, whereas non-zero $\hat\alpha_2$ induces a precession
of the orbital angular momentum about ${\bf w}$. The best current
limits come from observations of the binary pulsar systems PSRs
J1012+5307 and J1738+0333, both of which have short orbital periods,
$\sim 0.60$ and $\sim 0.35$~days respectively, and extremely low
orbital eccentricities $e \lapp 10^{-7}$ \cite{sw12}. Furthermore,
both of these pulsars have optically identified binary companions
which, together with proper motion measurements from the timing
observations, allow the three-dimensional space velocity of the binary
system to be determined. For these pulsars, the observational data
span is sufficiently long ($\gapp 10$~years) that relativistic
periastron advance has significantly changed the orientation of the
intrinsic eccentricity vector relative to the direction of any forced
eccentricity, resulting in a potentially detectable change in the
total eccentricity. The strongest limit on $\hat\alpha_1$ comes from
observations of PSR J1738+0333 as shown in Figure~\ref{fg:alpha1},
conservatively $\hat\alpha_1 < 4 \times 10^{-5}$. This is not only
better than the best weak-field limit of $\alpha_1 < 2 \times 10^{-4}$
from lunar laser ranging \cite{mwt08} but also constrains
strong-field effects as well.

A precession of the orbital axis about the system velocity vector
${\bf w}$ induced by a non-zero value of $\hat\alpha_2$ is potentially
observable as a time variation $\dot x$ in the projected semi-major
axis of the pulsar orbit $x \equiv a_1 \sin i$. $\dot x$ is one of the
possible post-Keplerian parameters in standard timing solutions and
significant values have been measured for both PSRs J1012+5307 and
J1738+0333 \cite{sw12}. There are several possible contributions to
the observed $\dot x$ but in Ref.~\refcite{sw12} it is shown that all
of these are negligible in these systems except that due to the
changing orbit inclination $i$ resulting from proper motion of the
system. This is a function of the unknown $\Omega$ values and for
certain $\Omega$ values there is no constraint. Consequently only a
probabilistic limit on $\hat\alpha_2$ can be obtained. By combining
results for the two pulsars, a 95\% confidence limit of
$|\hat\alpha_2| < 1.8\times 10^{-4}$ is obtained.

This is not as constraining as a solar-system limit $|\alpha_2| <
2.4\times 10^{-7}$ obtained from the present deviation of the Sun's
spin axis from the solar-system orbit normal \cite{nor87b}, a limit
that rests on the assumption that the two axes were aligned at the
time of formation of the solar system.

\begin{figure}[ht]
\centerline{\includegraphics[width=85mm]{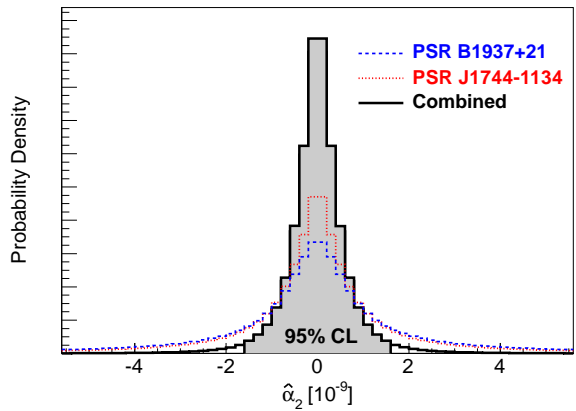}}
\caption{Constraints on the strong-field PPN parameter $\hat\alpha_2$
  based on the long-term stability of the mean pulse profiles for the
  isolated MSPs B1937+21 and J1744$-$1134.
  (Ref.~\protect\refcite{sck+13}) }
\label{fg:alpha2}
\end{figure}

However, pulsars provide an even stronger constraint based on the
stability of the spin axis of isolated pulsars. Any precession of the
spin axis of a pulsar is likely to result in changes in the observed
pulse profile (as observed with geodetic precession in binary pulsars
as discussed in \S\ref{sec:rel_grav} above). Shao et al.\cite{sck+13} compared
mean pulse profiles for the isolated MSPs B1937+21 and J1744$-$1134
taken 10 -- 12 years apart with the same observing system and found no
perceptible change in the pulse width at 50\% of the peak
amplitude. They interpreted these results by assuming a circular beam
profile for the main pulse in PSR B1937+21 and for PSR
J1744$-$1134. All of the angles in the problem can be estimated from
modelling of radio and gamma-ray observations of the pulsar, taken
together with known direction of the system velocity ${\bf w}$ with
respect to the cosmic microwave background, except the angle of the
projected spin axis on the sky. Probability histograms for
$\hat\alpha_2$ allowing for this unknown angle are shown in
Figure~\ref{fg:alpha2}. The final result for the 95\% confidence upper
limit is $|\hat\alpha_2| < 1.6\times 10^{-9}$, about four orders of
magnitude better than the limit described above based on orbital
precession in pulsar binary systems and two orders of magnitude better
than the limit based on solar spin precession. The assumption of
circular beams in these MSPs is problematic, since there is good
evidence for caustic enhancement in MSP pulse profiles, both radio and
gamma-ray\cite{rmh10,gjv+12}, which would tend to elongate the
beam in the latitude direction, reducing the effect of precession on
the observed pulse profiles. However, taking this into account would
probably increase the limit by a factor less than ten, and so this
limit would remain the best available. 

The stable pulse profiles of PSRs B1937+21 and J1744$-$1134 have also
been used to limit the PPN parameter $\xi$ describing local position
invariance (LPI), also known as the Whitehead parameter, and its
strong-field counterpart $\hat\xi$ (Ref.~\refcite{sw13}). The centripetal
acceleration of Galactic rotation results in an anisotropy in the
local gravitational field at a pulsar resulting in a precession of the
pulsar spin vector around the direction of the Galactic acceleration
with period.
\begin{equation}
\Omega_p = \xi \left(\frac{2\pi}{P}\right)\left(\frac{v_G}{c}\right)^2
\cos\psi
\end{equation}
where $v_G$ is the velocity of Galactic rotation at the pulsar and
$\psi$ is the (unknown) angle between the pulsar spin vector and the
Galactic acceleration. Combining the results for the two pulsars in a
way analogous to that for the $\alpha_2$ test described above, Shao and
Wex \cite{sw13} obtain a 95\% confidence limit $|\hat\xi| < 3.9\times
10^{-9}$. Even with the qualification about beam shapes mentioned
above, this limit is at least two orders of magnitude better than the
next best limit obtained from considering of the evolution of the
solar spin-misalignment angle over the lifetime of the solar system
\cite{sw13}.

The PPN parameter $\zeta_2$ is one of a number of parameters that may
be non-zero in gravitational theories that violate conservation of
total momentum \cite{wil14}. A non-zero $\zeta_2$ would result in an
acceleration of the centre of mass of a binary system that changes
direction with periastron precession. This is best measured by looking
for a change in the rate of orbital decay in a binary system with large periastron
precession and a long data span.  Will \cite{wil92} used observations of
PSR B1913+16 with a 15-year data span to obtain a limit $|\zeta_2| < 4\times
10^{-5}$. Obviously, this test could be made more stringent using the 
long data spans now available for PSR B1913+16 and other double-neutron-star 
systems with high $\dot\omega$. 

\subsubsection{Dipolar gravitational waves and the constancy of G}\label{sec:dipoleGW}
As described above, because of its tiny eccentricity, the pulsar --
white-dwarf binary system PSR J1738+0333 has played a
significant role in limiting the PPN parameters describing preferred
frame effects. This system is composed of two very different stars,
the neutron star we see as a pulsar and the companion which we know to
be a white dwarf of mass 0.181~M$_\odot$ through its optical
identification \cite{avk+12}. This and its short orbital period ($\sim
8.5$~h) make it an ideal system for testing theories of gravity that
predict a dipolar component to gravitational-wave damping as well as
general scalar-tensor theories \cite{fwe+12}.

From analysis of about 10 years of timing data obtained at Parkes and
Arecibo observatories, Freire et al.\cite{fwe+12} find an observed
rate of orbital decay for PSR J1738+0333 of $\dot P_b = (-17.0\pm
3.1)\times 10^{-15}$. To obtain the intrinsic rate of orbital decay,
kinematic contributions from the differential acceleration of the
binary system and the solar system in the Galactic gravitational field
and the Shklovskii effect due to transverse motion must be subtracted,
giving $\dot P_b^{\rm Int} = (-25.9 \pm 3.2)\times 10^{-15}$. Since
the orbital parameters and the masses of the two stars are well known,
the orbit decay due to GR can be accurately determined: $\dot P_b^{\rm
  GR} = (-27.7^{+1.5}_{-1.9})\times 10^{-15}$, leaving a residual
orbit decay of $\dot P_b^{\rm Res} = (2.0^{+3.7}_{-3.6})\times
10^{-15}$.

This residual orbit decay is consistent with zero, which can be
interpreted as a further confirmation of the accuracy of GR. However,
because of the very different nature of the two stars in this binary
system, this result also places strong constraints on theories of
gravity that predict a dipolar component to gravitational-wave
emission. Besides a dipolar component $\dot P_b^{\rm D}$, there are
several other possible contributions to $\dot P_b^{\rm Int}$:
\begin{equation}
\dot P_b^{\rm Int} = \dot P_b^{\dot M} + \dot P_b^{\rm T} + 
\dot P_b^{\rm D} + \dot P_b^{\dot G}
\end{equation}
where $\dot P_b^{\dot M}$ is due to mass loss from the binary system,
$\dot P_b^{\rm T}$ is a term resulting from tidal effects on the white
dwarf (tidal effects on the neutron star are negligible) and $\dot
P_b^{\dot G}$ is decay resulting from a possible variation in the
gravitational `constant' $G$ \cite{dt91}. Freire et al.\cite{fwe+12}
show that the likely $\dot M$ and tidal terms are small for this
system, $\lapp 10^{-15}$, and so the limit on $\dot P_b^{\rm Int}$ is
effectively a limit on $\dot P_b^{\rm D} + \dot P_b^{\dot G}$.

Within certain restrictions on strong-field effects \cite{fwe+12},
the dipole term is given by:
\begin{equation}\label{eq:dipole-gw}
\dot P_b^{\rm D} = -\frac{4\pi^2}{P_b} T_\odot m_c \frac{q}{q+1}
\kappa_D {\mathcal S}^2 + {\mathcal O}(s^3)
\end{equation}
where $q = m_1/m_2$ is the mass ratio (with subscript 1 refering to
the neutron star), ${\mathcal S} = s_1 - s_2$ is the difference in
``sensitivity'' $s$ of the mass of each body to a scalar field $\phi$, where
\begin{equation}
s \equiv \left(\frac{d\ln m(\phi)}{d\ln\phi}\right),
\end{equation}
and $\kappa_D$ is a body-independent constant that describes the
dipole self-gravity contribution in a given theory of gravity
(see, e.g., Ref.~\refcite{wil14}). The sensitivity $s_i$ depends on the
stellar equation of state and, for neutron stars, is typically about
0.15, whereas for a white dwarf it is $\sim 10^{-4}$. Therefore if
$\kappa_D$ is non-zero, dipole radiation will contribute to the orbit
decay. 

The remaining term in the residual orbit decay is that due to possible
variations in $G$. In weak gravity, $\dot G/G$ has been constrained to
be less than $4\times 10^{-13}$~yr$^{-1}$ from lunar laser ranging
experiments \cite{hmb10}, giving
\begin{equation}\label{eq:pb_gdot}
\dot P_b^{\dot G} = -2\frac{\dot G}{G}P_b < 0.8\times 10^{-15}.
\end{equation}
and hence $|\kappa_D| < 2\times 10^{-4}$. However, it is also possible
to obtain independent estimates of the effects of dipole radiation and
$\dot G$ by combining the J1738+0333 results with those for PSR
J0437$-$4715 \cite{dvtb08}. This southern pulsar is in a wider orbit
than PSR J1738+0333 and hence has a different mix of the dipole and
$\dot G$ components, allowing them to be separated. After accounting
for the fact that a changing $G$ will also change the stellar masses
\cite{nor90}, the formal results are $\dot G/G = (-0.6\pm 1.6)\times
10^{-12}$~yr$^{-1}$ and $\kappa_D = (-0.3\pm 2.0)\times 10^{-4}$, both
effectively upper limits. While the limit on $\kappa_D$ is the best
available, the derived limit on $\dot G/G$ is about an order of
magnitude weaker than the result (actually a limit on the variation of
$GM_\odot$) from the Mars Reconnaissance Orbiter \cite{kaf+11} and
from lunar laser ranging \cite{hmb10}.

Interestingly, pulsars have provided two other independent limits on $\dot
G/G$. Thorsett\cite{tho96a} used determinations of neutron star masses from
timing observations of double-neutron-star systems that formed many
gigayears ago. In standard formation scenarios, the mass of a neutron
star depends on the Chandrasekhar mass, the maximum possible mass of a
white dwarf star, just prior to the collapse to a neutron star. The
Chandrasekhar mass is proportional to $G^{-3/2}$ and so the observed
small range of neutron star masses implies that $\dot G/G < 4\times
10^{-12}$. This limit has been somewhat weakened by recent
discoveries of both less massive and more massive neutron stars in
pulsar binary systems (see Ref.~\refcite{kkdt13} for a recent review). 

The very small observed rate of change of pulsar
period $\dot P$ observed in some pulsars (after correction for
kinematic effects) may be used to set a further
independent limit \cite{wdk+00}. A variation in $G$ will result in an inverse
variation in the stellar moment of inertia, with the exact relation
depending on the neutron-star structure. If the observed (intrinsic)
$\dot P$ is entirely attributed to this effect, a
limit of $\dot G/G \lapp 2\times 10^{-11}$ is obtained. 

\subsubsection{General scalar-tensor and scalar-vector-tensor theories}
Many alternate theories of gravity can be expressed in a
``tensor-scalar'' framework in which a scalar field $\phi$ contributes
to the ``physical metric'' $\tilde g_{\mu\nu}$ through a coupling
function $A(\phi)$:
\begin{equation}\label{eq:t-s}
\tilde g_{\mu\nu} \equiv A^2(\phi) g_{\mu\nu}
\end{equation}
where $g_{\mu\nu}$ is the usual tensor metric. The coupling constant
may be expressed in different ways \cite{wil14}, one of which is as
an expansion around the asymptotic value of the scalar field $\phi_0$:
\begin{equation}\label{eq:scalar}
\ln A(\phi) = \ln A(\phi_0) + \alpha_0(\phi - \phi_0) + 
   \beta_0(\phi - \phi_0)^2 + ...
\end{equation}
(Ref.~\refcite{de92}). For GR, $\alpha_0 = \beta_0 = 0$. In the
well-known example of a tensor-scalar theory, the Brans-Dicke theory,
the scalar coupling is described by a single parameter
$\omega_{BD}$. For $\omega_{BD} \rightarrow \infty$, the Brans-Dicke
theory approaches GR. For this theory $\alpha_0 = 1/(2\omega_{BD}+3)$
and $\beta_0 = 0$. In other theories both the linear and quadratic
terms in Equation~\ref{eq:scalar} (and higher-order terms) may be
non-zero.

\begin{figure}[ht]
\centerline{\includegraphics[width=85mm]{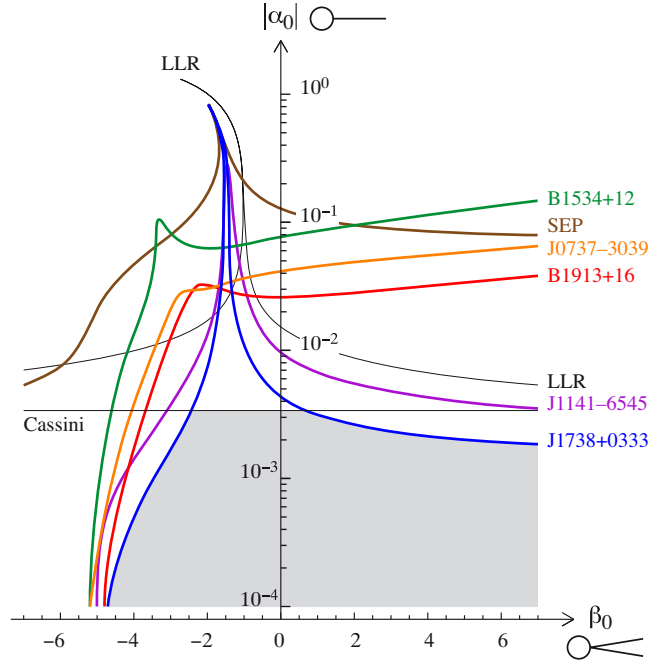}}
\caption{Constraints on the scalar-field parameters in the
  ($|\alpha_0|$,$\beta_0$) plane from various observational
  tests. Only the region below each line is allowed by the
  corresponding test. ``SEP'' refers to the test of the strong
  equivalence principle based on low-eccentricity pulsar --
  white-dwarf binary systems \cite{gsf+11}, ``LLR'' refers to lunar
  laser ranging results \cite{wtb12} and Cassini to the ``Shapiro
  delay'' experienced by signals to and from the Cassini spacecraft
  (on its way to Saturn) as its line of sight passed close to the Sun
  \cite{bit03}. Other binary-system tests are labelled according to
  the pulsar concerned. In GR, both $\alpha_0$ and $\beta_0$ are
  zero. (Ref.~\protect\refcite{fwe+12}) }
\label{fg:a0b0_ts}
\end{figure}

The various observational constraints on PPN and post-Keplerian
parameters can be expressed as limits in the ($\alpha_0$,$\beta_0$)
space for tensor-scalar theories as shown in Figure~\ref{fg:a0b0_ts}
\cite{fwe+12}. The Cassini experiment \cite{bit03} placed a strong
limit on the PPN parameter $\gamma_{PPN} = 1+(2.1\pm 2.3)\times
10^{-5}$ which translates to a limit on $|\alpha_0|$ of about
0.003. Only the limits on dipole gravitational radiation from the
asymmetric binary systems PSR J1141$-$6545 \cite{bbv08} and PSR
J1738+0333 \cite{fwe+12} rival the Cassini limit over most of the
space. Since the precision of these measurements will increase with
time, it seems likely that ultimately binary systems such as these will
provide the strongest constraints on tensor-scalar theories.

PSR J0348+0432 and its white-dwarf companion form another asymmetric
binary system, one that is distinguished by its very short orbital
period (2.46 hours) and massive neutron star
($2.01\pm0.04$~M$_\odot$).\cite{afw+13} The high neutron-star mass is
of interest since some scalar-tensor theories predict a strongly
non-linear relationship between the strength of dipole GW emission and
neutron-star mass or self-gravity (cf.,
Equation~\ref{eq:dipole-gw}). For most of the parameter space of the
class of theories discussed by Freire et al.\cite{fwe+12}, the limits
on effective scalar coupling from PSR J0348+0432 are currently not as
strong as those from PSR J1738+0333 but, because of the high
neutron-star mass, they place stronger limits on some other theories
with a greater degree of non-linear coupling.\cite{wex14}

Bekenstein\cite{bek04} has proposed a relativistic generalisation of the
so-called ``MOND'' theory of gravity \cite{mil83} that seeks to avoid
the need for dark matter in galactic dynamics. The generalisation
invokes an additional vector field and hence is known as a
tensor-vector-scalar theory. Such theories relax some of the
constraints on tensor-scalar theories, in particular, the dipole
radiation constraints, and allow significantly larger values of
$\alpha_0$. However, as shown by Freire et al.\cite{fwe+12}, the binary pulsar
results still significantly constrain theories of this type and in
fact are more constraining than solar-system tests. With
future observations, binary pulsar tests have the potential to
make this class of theories untenable. 

In another example of the use of pulsar observations, especially the
limits on dipolar-GW radiation, to place limits on gravitational
theories, Yagi et al.\cite{ybby14} have strongly limited the allowed parameter
space for the LLI-violating ``Einstein-{\AE}ther'' and the
``Khronometric'' theories.

\subsection{Future Prospects}
As described above, GR has provided an accurate description of all
pulsar timing results obtained so far. However, continued refinement
of existing methods and development of new tests is highly
desirable. Continued pulsar timing measurements, especially with the
advent of new and more sensitive observing facilities such as the
500-m Arecibo-type {\it FAST} radio telescope in China \cite{nlj+11}
and the Square Kilometre Array ({\it SKA}) in South Africa and
Australia \cite{cr04b} will certainly improve on existing limits and
enable new tests of gravitational theories. They may even demonstrate
a failure of GR and hence a need for a modified or conceptually
different theory of gravity. Conversely, if GR is assumed to be valid,
results of astrophysical significance can be deduced from the
observations. For example, as described in Section~\ref{sec:J0737},
observations of higher-order terms in the relativistic perturbations
may enable a measurement of the moment of inertia of a neutron star.

These more sensitive radio telescopes can also be used to search for
previously unknown pulsars and binary systems that are suitable for
tests of gravitational theories.  Past experience has shown that
pulsar searches repeatedly turn up new classes of object. This
potential is wonderfully illustrated by the recent discovery of the
pulsar triple system PSR J0337+1715 which promises to provide a strong
limit on violations of the Strong Equivalence Principle. A dream for
such searches is the discovery of a pulsar in a close orbit around a
black hole as this would offer much more stringent tests of gravity in
the strong-field regime\cite{lewk14}. Discovery of a pulsar orbiting
the black hole at the centre of our Galaxy with an orbital period of a
few months or less could even allow a test of the so-called
``no-hair'' theorem for black holes\cite{lwk+12}.

\section{The Quest for Gravitational-Wave Detection}\label{sec:gw_detn}
The direct detection of the gravitational waves (GWs) predicted by
Einstein's general theory of relativity and other relativistic
theories of gravity is one of the major goals of current
astrophysics. As described in \S\ref{sec:rel_grav} above, we have
excellent evidence from the orbital decay of binary systems for the
existence of GWs at the level predicted by GR, but up to now there
has been no direct detection of the changing curvature of spacetime
induced by a passing GW. This changing curvature induces a change in
the proper distance between two test masses, described by the
gravitational strain $h = \delta L/L$. The problem is that, for any
likely source, $h$ is tiny. For example, the {\it LIGO} gravitational-wave
detector \cite{aaa+09j} hopes to detect the merger of two neutron
stars at a distance of 100 Mpc for which $h\sim 10^{-22}$, a change in
the length of its 4-km arms of $10^{-18}$~m or $10^{-3}$ of the
diameter of a proton! 

Pulsar timing can measure a change in the proper distance between the
pulsar and the telescope. Systematic changes in timing residuals for a
given pulsar reflect unmodelled changes in the effective time of
emission, the pulsar position, the propagation path or the position of
the telescope. With care, and with observations of multiple pulsars,
residual delays due to changing proper distances can be isolated,
effectively giving us a set of interferometers with baselines of
$\gapp 10^{16}$~km! However, even in the best cases, we can only
measure the interferometer ``phase'' to about 100~ns, so that the
limiting strain is about $10^{-18}$. Unlike ground-based
laser-interferometer systems, which are most sensitive to GW signals
with frequencies around 100 Hz, pulsar timing systems are most
sensitive to signals with frequencies around the inverse of the data
span, typically a few nanohertz. Potential sources of detectable GWs in this
low-frequency band include super-massive black-hole (SMBH) binary
systems in distant galaxies and cosmic strings in the early Universe. 

\subsection{Pulsar Timing Arrays}
\begin{figure}[ht]
\centerline{\includegraphics[width=85mm]{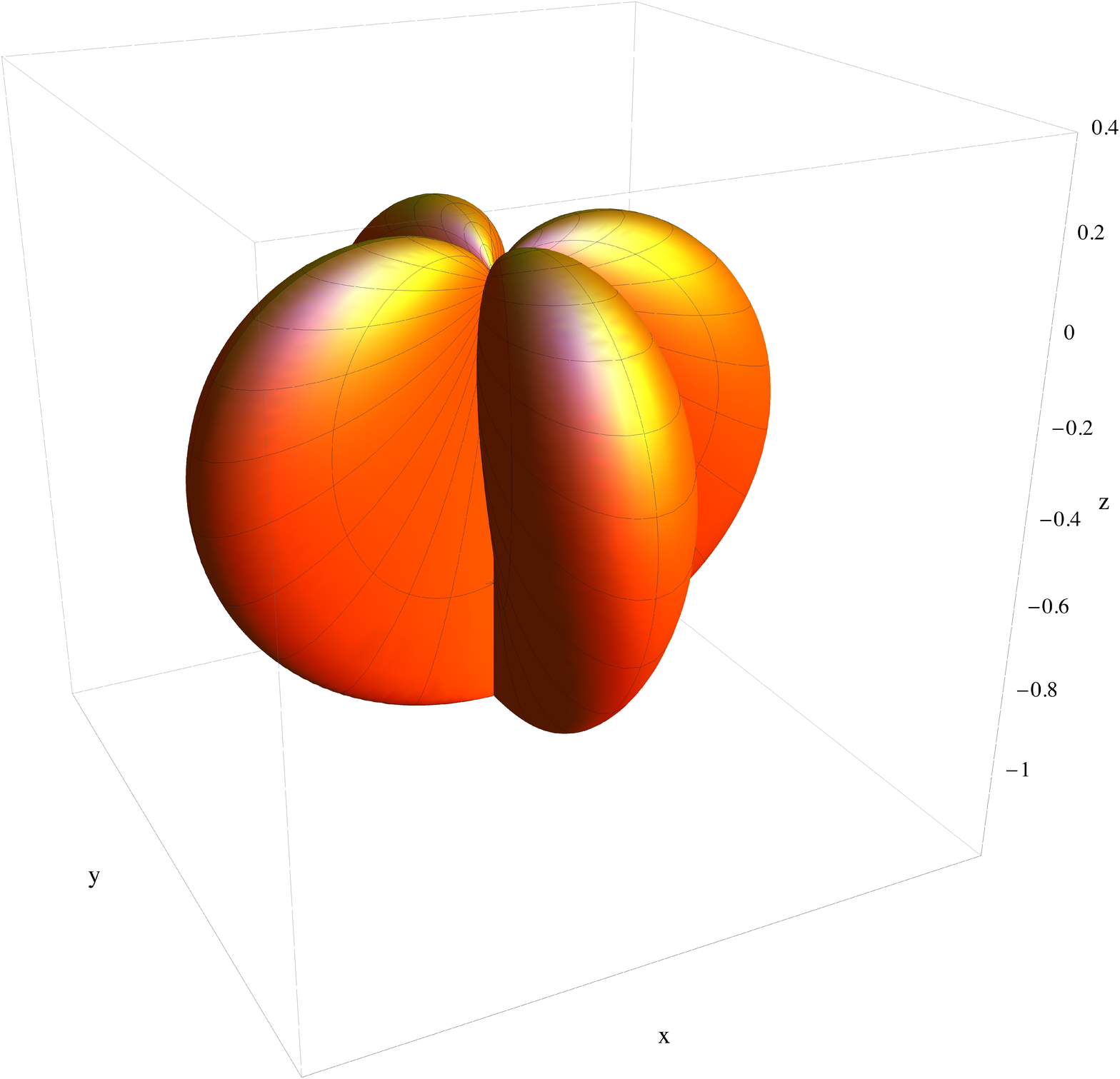}}
\caption{Effective ``antenna pattern'' for detection of a GW with
  pulsar timing. The wave is propagating in the $+z$ direction and is
  assumed to have the + polarisation. The pattern for the $\times$
  polarisation is the same but rotated by $45\degr$ about the $z$
  axis.  (Ref.~\protect\refcite{cs12})}
\label{fg:gw_ant}
\end{figure}

The effect of GWs on pulsar timing signals was first considered by
Sazhin\cite{saz78} and Detweiler\cite{det79}, with the latter being the
first to consider the effect of a GW from a distant source passing
over a pulsar and the Earth. For this case, it can be shown that the
net effect on the observed pulsar arrival times is simply the
difference between the effect of the GW passing over the pulsar and
the effect of the GW passing over the Earth. For a GW travelling in
the $\hat z$ direction, the redshift $z$ of the pulse frequency $\nu$
for a pulsar at distance $d$ with direction cosines ($\alpha$,
$\beta$, $\gamma$) is given by:
\begin{equation}\label{eq:gwz}
z(t) = \frac{\nu_0 - \nu(t)}{\nu_0} = \frac{\alpha^2-\beta^2}
  {2(1+\gamma)} \Delta h_+ + \frac{\alpha\beta}{1+\gamma} \Delta h_\times 
\end{equation}
where $\Delta h_A = h^p_A -h^E_A$, with $A$ representing the two
possible wave polarisation states ($+$,$\times$) in GR, and $h^p_A(t-d/c)$
and $h^E_A(t)$ the gravitational strain at the pulsar and the Earth,
respectively.\footnote{See Ref.~\refcite{abc+09} for a rederivation of the
  Detweiler result.} The observed timing residuals are then given
by the integral of the redshift,
\begin{equation}
R(t) = \int_0^t z(t^\prime) dt^\prime.
\end{equation}
The coefficients multiplying the $\Delta h_+$ and $\Delta h_\times$ terms
are the ``antenna patterns'' for the two polarisations as illustrated
in Figure~\ref{fg:gw_ant}. A pulsar located in the GW
propagation direction has zero sensitivity to the GW since its
direction cosines $\alpha$ and $\beta$ are zero. Furthermore, despite
the $(1+\gamma)$ term in the denominator of Equation~\ref{eq:gwz}, the
response for a pulsar exactly in the $-z$ direction (the same direction as the
GW source) is also zero. This comes about because of a cancellation of
the $(1+\gamma)$ term by the expansion of $\Delta h_A$ when $1+\gamma
\ll 1$ (cf. Ref.~\refcite{lwk+11}). 

A pulsar timing array (PTA) consists of a set of pulsars spread across
the sky which have precise timing measurements over a long data
span. The detection of GWs by PTAs depends on the correlated timing
residuals for different pulsars given by the Earth term $h^E_A(t)$ in
Equation~\ref{eq:gwz}. GWs passing over the pulsars produce
uncorrelated residuals because of both the retarded time and the
different GW environment for each pulsar. Also the pulsars themselves
have uncorrelated timing noise at some level, either intrinsic or
resulting from uncorrected variations in interstellar delays. Because
the expected GW strain is so weak, only MSPs have sufficient timing
precision to make GW detection with PTAs feasible.

For an isolated source of continuous GWs, say an SMBH binary system in
a nearby galaxy, in principle, both the Earth term and and the pulsar
term in Equation~\ref{eq:gwz} could be detected. For a rapidly
evolving source, the pulsar term and the Earth term may be at
different frequencies because of the retarded time of the pulsar term
(see, e.g., Ref.~\refcite{jllw04}). They can then be added incoherently to
increase the detection sensitivity. For a non-evolving source, i.e.,
$\delta f_b \ll 1/T$ where $\delta f_b$ is the change in binary
orbital frequency over the pulsar timing data span $T$, in principle
the pulsar term and the Earth term could be summed coherently for
optimal sensitivity. As is discussed further in
Section~\ref{sec:future} below, unfortunately we currently don't know
enough pulsar distances to sufficient accuracy to make this
coherent addition possible. In a PTA, the pulsar terms therefore add
with random phase, washing out the fringes in the antenna pattern
(see also Ref.~\refcite{lwk+11}) and adding ``self-noise'' to the signal from
the Earth term. Since the antenna pattern (Figure~\ref{fg:gw_ant}) has
a maximum for pulsars roughly in the same direction as the GW source, the
maximum response of a PTA is toward the greatest concentration of
pulsars in the array.

\begin{figure}[ht]
\centerline{\includegraphics[width=85mm,angle=270]{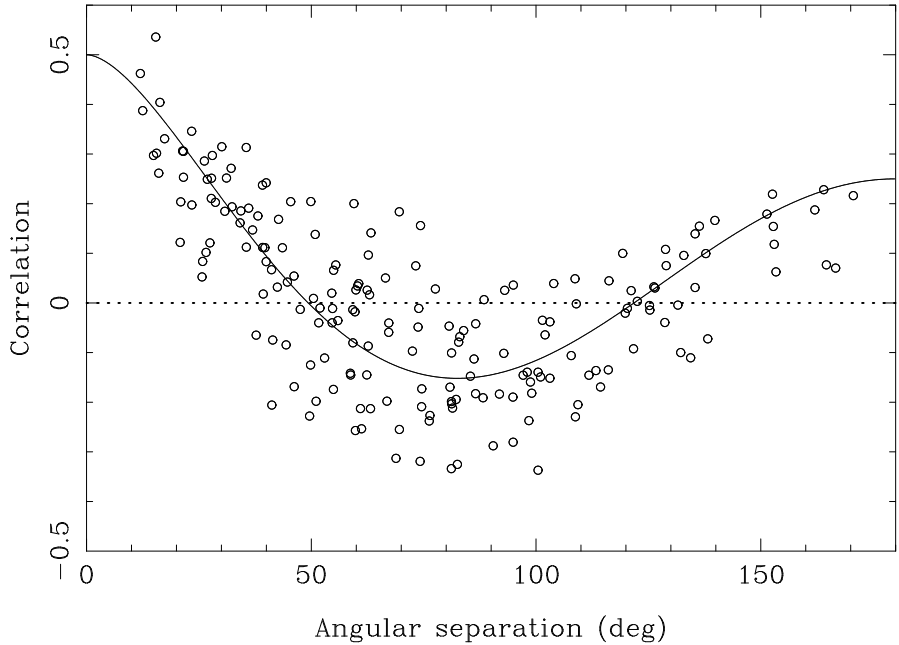}}
\caption{The Hellings \& Downs correlation function, i.e., the
  correlation between timing residuals for pairs of pulsars as a
  function of their angular separation for an isotropic stochastic
  background of GWs. Also shown are simulated correlations between the
  20 pulsars of the Parkes PTA for a single realisation of a strong GW
  signal that dominates all other noise
  contributions. (Ref.~\protect\refcite{hjl+09})}
\label{fg:hd}
\end{figure}

A stochastic background of nanohertz GW from many SMBH binary systems
in distant galaxies is likely to be the signal first detected by
PTAs. To a first approximation, this background is also likely to be
statistically isotropic, i.e., the expectation value $\langle
h^2\rangle$ is independent of direction when averaged over typical
data spans. Hellings and Downs\cite{hd83} were the first to show that,
in this case, the correlation between GW-induced timing residuals for
two pulsars separated by an angle $\theta$ on the sky is dependent
only on $\theta$ and not on the sky positions of the two pulsars. The
zero-lag correlation function, commonly known as the Hellings \& Downs
curve and obtained by integrating the product of the antenna patterns
(Figure~\ref{fg:gw_ant}) for the two pulsars over all possible GW
propagation directions, is given by:
\begin{equation}\label{eq:hd}
c_{\rm HD} = \frac{1}{2} + \frac{3x}{2} \left(\ln x - \frac{1}{6}\right),
\end{equation}
where $x = (1 - \cos\theta)/2$, and is plotted in
Figure~\ref{fg:hd}. $c_{\rm HD}$ goes negative for angular separations
around $90\degr$ and then positive again for pulsars that are
more-or-less opposite on the sky -- this is a direct consequence of
the quadrupolar nature of GWs. It is also important to note that the
limiting value as $\theta \rightarrow 0$ is 0.5, not 1.0. This is a
consequence of the fact that the pulsar terms in Equation~\ref{eq:gwz}
are uncorrelated and, on average, of equal amplitude to the Earth
term. The scatter in the simulated correlations results from the
random phases of the pulsar terms and illustrates the ``self-noise''
that limits the sensitivity of PTA experiments in the strong-signal
limit. 

\subsection{Nanohertz Gravitational-Wave Sources}\label{sec:gwsources}

\subsubsection{Massive black-hole binary systems}\label{sec:gwb}
There is good evidence that massive black holes form in the centre of
galaxies at very early times (see, e.g.,
Ref.~\refcite{vfs+13}) and also that merger events play a
major role in galaxy growth (see, e.g., Ref.~\refcite{sbbw09}). When
two galaxies, each containing a central massive black hole, merge,
dynamical friction will result in the two black holes migrating to the
centre of the merged galaxy to form a binary system, with an estimated
timescale for the migration of the order of giga-years (see, e.g.,
Ref.~\refcite{kbb+12}). When the binary separation is less than about
1~pc, loss of energy to GWs becomes the dominant orbital decay
mechanism and the binary system will ultimately coalesce, becoming a
strong GW source as it spirals in. There remains much controversy
about the efficiency of orbital decay mechanisms as the binary
separation approaches a parsec -- known as the ``last parsec
problem''. Some (see, e.g., Ref.~\refcite{khbj13,rws+14})
argue that dissipation mechanisms will quickly move the binary system
through this phase, whereas others (see, e.g., Ref.~\refcite{cmt13})
argue that the binary is likely to stall at separations where the
gravitational decay is ineffective. Detection of a stochastic GW
background (GWB) would resolve this issue.

Since large numbers of binary systems with different orbital periods
contribute to the GWB, it is a broadband signal which is best described in
the spectral domain. It is convenient to express the amplitude of the
GW signal in terms of the dimensionless ``characteristic strain'', defined by
\begin{equation}
h_c = 2f|\tilde h(f)|
\end{equation}
where $\tilde h(f)$ is the Fourier transform of $h(t)$ and $f$ is the GW
frequency. (Note that, for a circular binary system, the frequency of
the emitted GW is twice the binary orbital frequency $f=2f_b$.) Two
other quantities that are often used to parameterise GW spectra and
detector spectral sensitivities are the square root of the one-sided
strain power spectral density
\begin{equation}
S_h^{1/2}(f) = h_c f^{-1/2}
\end{equation}
and the GW energy density as a fraction of the closure energy density
of the Universe
\begin{equation}\label{eq:omgw}
\Omega_{GW} = \frac{2\pi^2}{H_0^2} f^2 h_c^2(f)
\end{equation}
where $H_0$ is the Hubble constant. 

In order to understand the astrophysical implications of results
obtained from PTA experiments, it is necessary to have estimates of
the likely strength of signals from potential sources of nanohertz
GW. For a cosmological population of SMBH binary
systems at luminosity distance $D_L$ and redshift $z$, the local
energy density in GW at frequency $f$ is given by:
\begin{equation}
fS_E(f) = \int_0^\infty dz\int_0^\infty dM_c
\frac{d^2n}{dz dM_c} \frac{1}{(1+z)} \frac{1}{D_L^2} \frac{dE_g}{d\ln f_r}
\end{equation}
where $M_c = (M_1 M_2)^{3/5} (M_1 + M_2)^{-1/5}$ is the binary
chirp mass and $M_1$ and $M_2$ are the masses of the binary
components, $d^2n/(dz dM_c)$ is the comoving density of binary
systems with redshift and chirp mass between $z$ and $z+dz$ and
$M_c$ and $M_c + dM_c$, respectively, and
$dE_g/d\ln f_r$ is the total energy emitted by a single binary system
in the logarithmic frequency interval $d\ln f_r$, where $f_r = f(1+z)$
\cite{phi01,svc08,mcb14}.  The local GW energy density is related to
the local characteristic strain by
\begin{equation}
fS_E(f) = \frac{\pi c^2}{4G} f^2 h_c^2(f)
\end{equation}
and for a circular binary system
\begin{equation}
\frac{dE_g}{d\ln f_r} = \frac{G^{2/3}\pi^{2/3}}{3} M_c^{5/3} f_r^{2/3}.
\end{equation}
Therefore we have 
\begin{equation}\label{eq:gwb}
h_c^2(f) = \frac{4G^{5/3}}{3\pi^{1/3}c^2} f^{-4/3} \int_0^\infty dz
\int_0^\infty dM_c \frac{d^2n}{dz dM_c}
\frac{1}{(1+z)^{1/3}} \frac{1}{D_L^2} M_c^{5/3}.
\end{equation}

As Phinney\cite{phi01} has emphasised, the result that $h_c \propto
f^{-2/3}$ for a cosmological population of circular binary systems
decaying through GW emission is quite general and independent of any
particular cosmology, black-hole mass function or galaxy merger
scenario. Consequently the spectrum of the GWB is often parameterised
as follows:
\begin{equation}\label{eq:gwb_spec}
h_c(f) = A_{1{\rm yr}} \left(\frac{f}{f_{1{\rm yr}}}\right)^\alpha
\end{equation}
where $f_{1{\rm yr}} = (1{\rm yr})^{-1}$, $A_{1{\rm yr}}$ is the
characteristic strain at $f_{1{\rm yr}}$ and $\alpha = -2/3$ for the
case described above. For pulsar timing experiments, the one-sided
power spectrum of the timing residuals is given by
\begin{equation}\label{eq:gwb_res}
P(f) = \frac{1}{12\pi^2}\frac{1}{f^3} h_c^2(f).
\end{equation}
Consequently, a GWB produces a very ``red'' modulation of the timing
residuals with a spectral index of $-13/3$ for $\alpha = -2/3$.

\begin{figure}[ht]
\begin{minipage}{125mm}
\begin{tabular}{cc}
\mbox{\includegraphics[height=58mm]{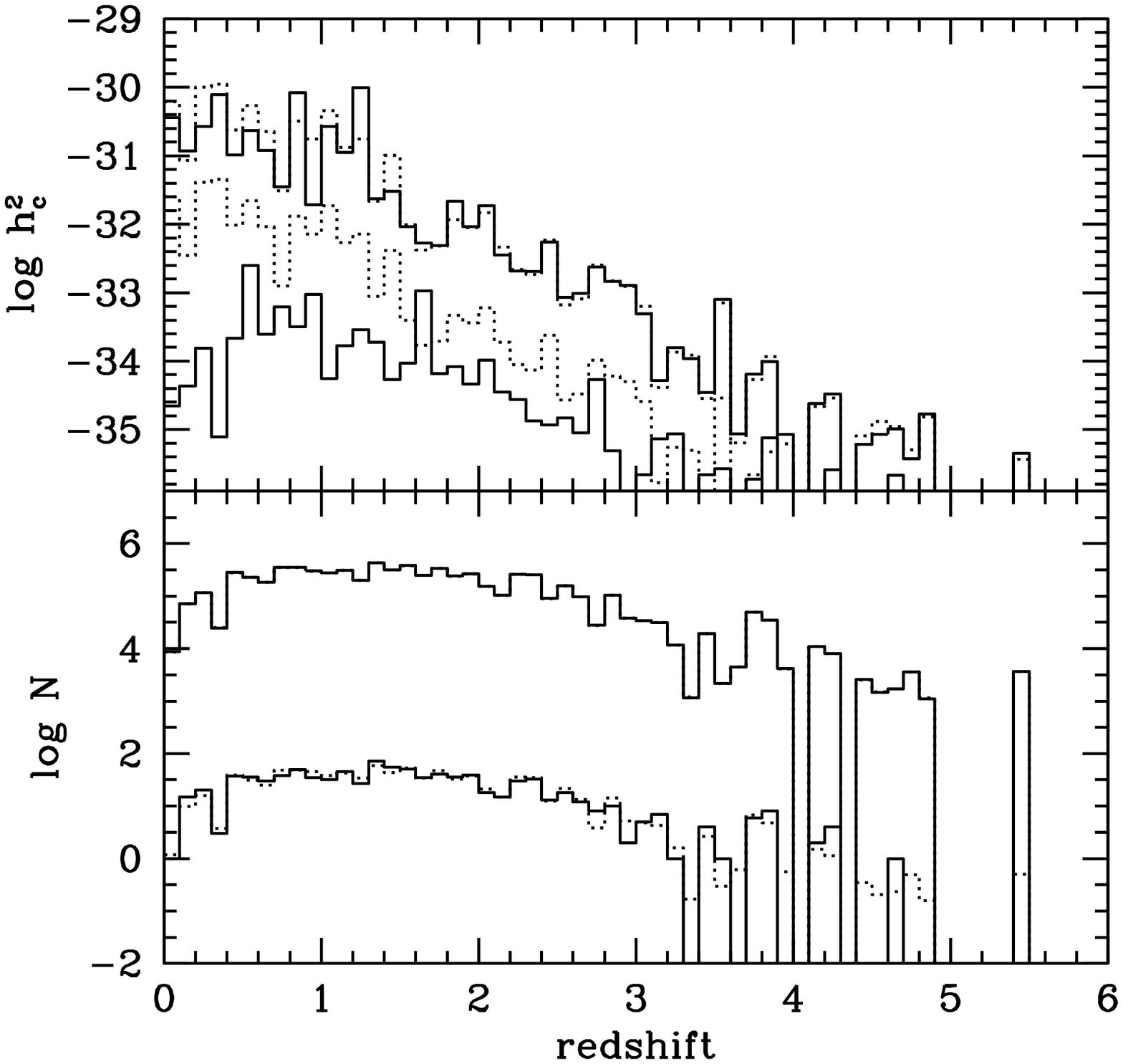}} &
\mbox{\includegraphics[height=58mm]{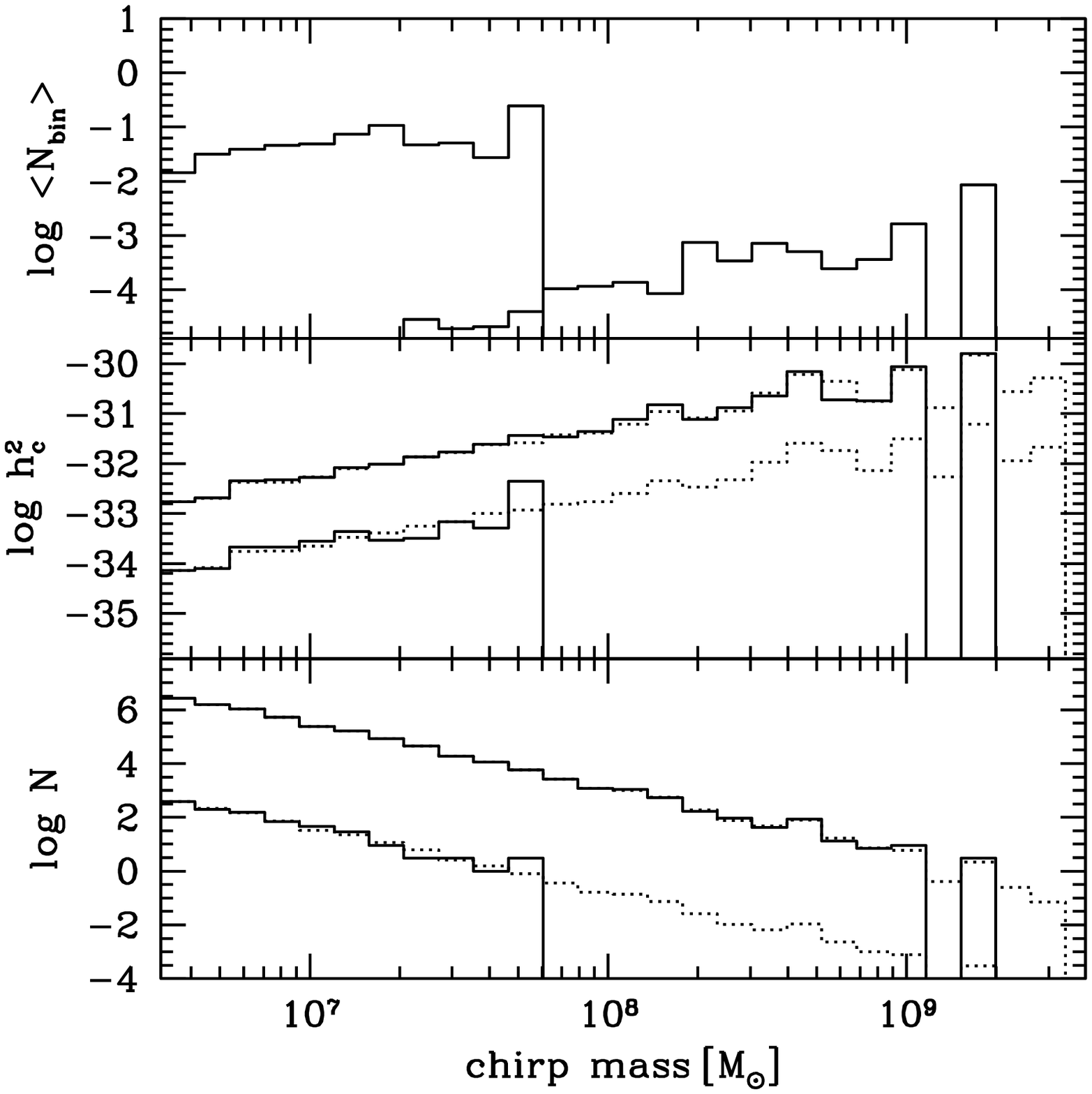}}
\end{tabular}
\caption{Left panel: Number of binary black-hole systems and their
  contribution to the characteristic strain $h_c$ of the GWB as
  function of source redshift $z$. The solid lines are results from a
  Monte Carlo approach and the dotted lines are from a semi-analytic
  analysis. In both panels, the upper histograms are for a GW
  frequency $f=8\times 10^{-9}$~Hz and the lower histograms for
  $f=10^{-7}$~Hz. Right panel: The lower two panels are as for the
  left panel but as a function of source chirp mass $M_c$. The upper
  panel shows the number of frequency bins spanned by the chirp over
  the 5-year span of the simulation. (Ref.~\protect\refcite{svc08})}
\label{fg:mbhb}
\end{minipage}
\end{figure}

\begin{figure}[ht]
\centerline{\includegraphics[width=85mm]{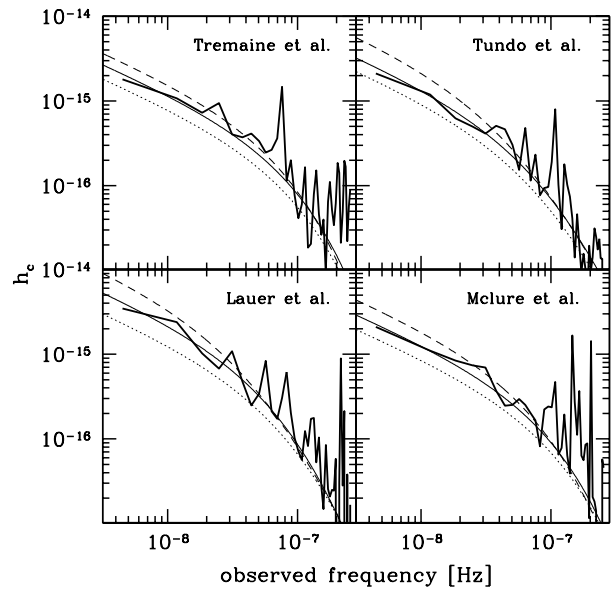}}
\caption{Characteristic strain spectrum for the GWB based on different
  prescriptions for black-hole growth by accretion between mergers
  (solid, dashed and dotted lines) and different black-hole mass
  functions (different panels). (Ref.~\protect\refcite{svv09})}
\label{fg:gwb_spec}
\end{figure}

In order to estimate the likely strength of this modulation, the
factor $d^2n/(dz dM_c)$ in Equation~\ref{eq:gwb} must be
evaluated. This requires a prescription for the cosmological evolution
of massive black-hole binary systems in galaxies. Different approaches
to this problem have been taken by different authors. An early paper
by Jaffe and Backer\cite{jb03} used observational constraints on close
galaxy pairs coupled with a black-hole mass function, whereas another
early paper by Wyithe and Loeb\cite{wl03a} used a prescription for
merger of dark-matter halos coupled with different scenarios for
growth of massive black holes in galaxies. The latter approach was
developed further by Sesana et al.\cite{svc08} who showed that the GWB
spectrum steepens at frequencies above about $10^{-8}$~Hz since the
number of binary systems contributing to the background at these
frequencies becomes small. This is illustrated in the left panel of
Figure~\ref{fg:mbhb} which shows that binary systems at $z \lapp 2$
contribute most of the strain to the GWB. The right panel shows that
massive binary systems with $M_c \gapp 10^8$~M$_\odot$ contribute most
of the low-frequency GWB but that only systems with $M_c \lapp
10^8$~M$_\odot$ contribute to the high-frequency end. Importantly, at
the high-mass end, only a few systems contribute to the GWB. As shown
in Figure~\ref{fg:gwb_spec}, this leads to a steepening and large
uncertainties in the expected GWB spectrum at frequencies $\gapp
10^{-8}$~Hz. These results were confirmed and extended by Sesana et
al.\cite{svv09} and Ravi et al.\cite{rwh+12} using the {\it Millenium}
simulation of the cosmological evolution of dark matter structures
\cite{swj+05} to define a merger history together with various
prescriptions for galaxy and black-hole formation and growth.

While there is some consensus about the form of the GWB spectrum,
there remain significant uncertainties. For example, Ravi et
al.\cite{rws+14} consider the effects of the stellar environment on
the late evolution of massive black-hole binary systems in the cores
of galaxies and conclude that the effect of dynamical friction is
important, both in extracting energy from the binary system and
inducing an eccentricity. Both of these have the effect of reducing
the strength of the predicted GWB, especially at the low-frequency
end, consequently making its detection by PTAs more difficult. On the
other hand, McWilliams et al.\cite{mop14} argue for a model in which
all black-hole growth is by merger rather than by accretion after
coalescence, which is the main contributor to black-hole growth in the
models discussed above. This leads to predictions of a significantly
larger GWB characteristic strain compared to previous predictions and,
hence, the imminent detection of the GWB by PTAs.

As Figure~\ref{fg:gwb_spec} indicates, there is a possibility that the
nearby universe could contain a massive black-hole binary system with
an orbital period of the order of a few years that actually dominates
the nanohertz GW spectrum. This opens up the exciting prospect of the
GW detection and study of an isolated supermassive black-hole binary
system using pulsar timing and even the possibility of detection and
study of the system in the electromagnetic bands -- so-called
``multi-messenger'' astronomy. Searches for binary GW sources will be
described in \S\ref{sec:pta} and the prospects for their detailed
study will be discussed in \S\ref{sec:future}.

\subsubsection{Cosmic strings and the early Universe}\label{sec:strings}
Cosmic strings and the related cosmic super-strings are
one-dimensional topological defects which may have formed in phase
transitions in the early Universe. Cosmic strings occur in standard
field-theory inflation models, whereas superstrings are found in brane
inflationary models. The idea that such strings will oscillate and
hence emit GWs was first proposed by Vilenkin\cite{vil81}. Such
oscillations may contribute to the stochastic GWB (see, e.g.,
Ref.~\refcite{ca92}) or generate bursts of GW radiation from string
cusps and kinks (see, e.g., Ref.~\refcite{dv01}). The amplitude of GWs
from cosmic strings is dependent on a large number of poorly known (or
unknown) parameters and hence is very uncertain (see, e.g.,
Ref.~\refcite{sbs12}). Key parameters are the string tension $\mu$,
usually parameterised by the dimensionless quantity $G\mu/c^2$, and
the size $\alpha$ of string loops relative to the horizon radius at
the time of birth. Other significant parameters for the GWB are the
intrinsic spectral index $q$ of the GW emission, a characteristic node
number $n_*$ for the high-frequency cutoff in the emission spectrum
and the probability $p$ of ``intercommutation'', that is, intersecting
strings dividing and the two parts exchanging. Such intercommutation
can, for example, form two smaller loops from an intersecting twist in
a larger loop. For standard strings, $p=1$ but it may be less for
superstrings.

\begin{figure}[ht]
\centerline{\includegraphics[width=85mm]{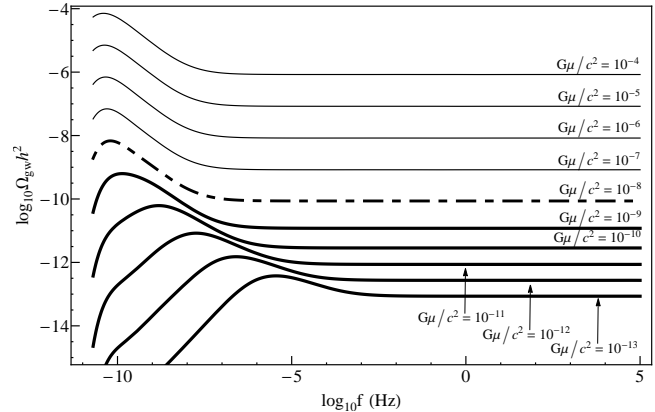}}
\caption{Energy density spectrum for GWs emitted by cosmic strings as
  a function of the dimensionless string tension $G\mu/c^2$. Other
  parameters are held fixed at values $\alpha=10^{-7}$, $q=4/3$,
  $n_*=1$ and $p=1$. The dashed spectrum is for the critical point
  where $\Gamma G\mu/c^2 \approx \alpha$; spectra above this are for
  large loops and spectra below are for small
  loops. (Ref.~\protect\refcite{sbs12})}
\label{fg:string_spec}
\end{figure}

Vibrating cosmic strings are likely to decay by emission of GWs in a
series of harmonics with fundamental frequency $2c/l$, where $l$ is
the length of the loop, with an initial value $\alpha D_H$, where
$D_H$ is the horizon distance at the time of loop creation. The rate
of energy loss for vibration mode $n$ of a given loop is:
\begin{equation}\label{eq:str_edot}
\frac{dE_{GW}}{dt}=\Gamma\frac{n^{-q}}{\sum_{m=1}^\infty m^{-q}} G\mu^2 c
\end{equation}
where $\Gamma$ is factor depending on the shape of the loop, typically
about 50 \cite{sbs12}. As the loop loses energy, it shrinks and
eventually disappears. The creation of loops through intercommutation
and their decay through GW emission sets up an equilibrium
distribution of loop sizes. Sanidas et al.\cite{sbs12} compute the
number density of loops as a function of loop length and time and
hence, using Equation~\ref{eq:str_edot}, the predicted spectrum of the
GWB from string loops as a function of the various parameters. As
Figure~\ref{fg:string_spec} shows, the spectrum is very broad,
extending all the way from nanohertz to Megahertz. Since the lowest
($n=1$) frequency is proportional to the string length, the weaker and
smaller loops do not contribute to the nanohertz background.

\subsubsection{Transient or Burst GW sources}\label{sec:gw_trans}
The prime target of ground-based laser-interferometer GW detectors is
the burst emitted at the coalescence of a double-neutron-star
system. Such a burst is intense for just a few milliseconds and
clearly cannot be detected by PTA experiments which typically are
sensitive to signals of duration between a few weeks and a few
years. Possible sources of longer-duration bursts are coalescence of
SMBH binary systems, highly eccentric massive
black-hole binary systems, and formation and decay of cusps and kinks
in cosmic strings. 

A major difference between detection of GW bursts
and continuous GW sources is that there is generally no interference
between the Earth term and the pulsar terms in the signal detected by
PTAs (Equation~\ref{eq:gwz}). This is because the duration of the
burst (by definition, less than the observational data span) is much
less than the light-time to the pulsars and so, except for a source in
the same direction as a pulsar, the burst will occur at very different
times in the Earth and pulsar terms. The pulsar term reflects the
effect of the burst on the pulsar at a time $d/c$ {\it before} the
burst arrives at the Earth, but the pulsar term is detected at a time
$(d/c)(1 + \cos\theta)$ {\it later} than the Earth term, where $d$ is
the pulsar distance and $\theta$ is the angle between the pulsar and
the GW propagation direction as seen from the Earth.

\begin{figure}[ht]
\centerline{\includegraphics[width=85mm]{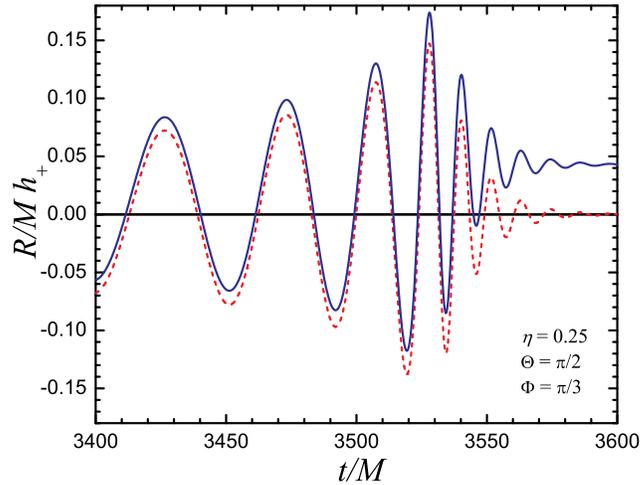}}
\caption{Gravitational waveform resulting from the coalescence of two
  equal-mass black holes (reduced mass ratio $\eta = 0.25$) with total
  mass $M$ at distance $R$. Both $M$ and $R$ are expressed in time
  units; the conversions to conventional units are $M \rightarrow M
  T_\odot$ and $R \rightarrow R/c$. $\Theta$ and $\Phi$ are assumed
  source directions. The dashed line is the predicted waveform if the
  gravitational memory effect is ignored. (Ref.~\protect\refcite{fav09})}
\label{fg:gw_wave}
\end{figure}

Figure~\ref{fg:gw_wave} shows the GW waveform produced by the
coalescence of two black holes in a coordinate system where all the
signal is in $h_+$. The maximum amplitude of the waveform is about 0.1
in the time units of Figure~\ref{fg:gw_wave}, or about $0.1 \; c M
T_\odot /D$ or $\sim 10^{-14} M_9/D_{Gpc}$ where $M_9 \equiv 10^{-9}
M$ is the total system mass in solar units and $D_{Gpc}$ is the
(comoving) distance in Gpc. Although this is comparable to the strain
sensitivities achieved by current PTAs for continuous GW signals (see
\S\ref{sec:pta} below), even for the largest SMBHs, the timescale of
the burst, $\sim 200 \; M T_\odot$ is only of order 10 days and the
period of the oscillation is about an order of magnitude less than
that. Not only is this too short to be resolved by any existing PTA,
but the sensitivity over this short interval would be much less than
that achieved for CW signals integrated over the entire data
span. Space-based laser interferometer systems such as the planned
{\it eLISA} \cite{aab+12} have the potential to directly detect these
bursts.

However, Figure~\ref{fg:gw_wave} also shows another effect known as
GW ``memory'' which is potentially detectable with
pulsar timing. During the coalescence event, a non-oscillatory
component to the gravitational strain builds up, so that at the end of
the ``ring-down'' phase, the strain has a permanent offset from the
pre-coalescence value.  The amplitude of the memory effect is 
\begin{equation}
h_m = \frac{\epsilon\eta MT_\odot}{D/c} \approx 10^{-15}\frac{M_9}{D_9}
\end{equation}
where $\epsilon \approx 0.07$ is the mass fraction contributing to the
memory effect and $\eta$ is the reduced mass fraction, 0.25 for
equal-mass binary components \cite{fav09,cj12}. 

This step change in $h$ produces a step change in the observed pulse
frequency $\Delta\nu/\nu = h_m$, i.e., a ``glitch''. This glitch
persists until it is reversed by the pulsar term at a time $(d/c)(1 +
\cos\theta)$ later. In a PTA, these reversals will occur at different
times for different pulsars. Of course, it is also possible that a GW
memory jump could be detected in the pulsar terms, but there it may be
confused with a real glitch in the intrinsic pulse frequency, whereas
in the Earth term there is a correlation in the effect on different
pulsars. However, glitches in MSPs are rare (only one very small
glitch detected so far: Ref.~\refcite{cb04}). Also, real pulsar glitches are
generally spin-ups, whereas a GW-memory jump may be of either sign,
depending on pulsar-source angle. Therefore, as Cordes and Jenet\cite{cj12}
have discussed, the pulsar terms may give an improved probability of
detection.

Black-hole binary systems with circular orbits emit GW at the second
harmonic of the orbital frequency, i.e., $f = 2f_b$. For eccentric
orbits, the GW emission becomes more burst-like as the accelerations
and hence GW power are greatest around periastron when the two black
holes are closest together \cite{pm63}. In the spectral domain, power
spreads to higher harmonics and also to the fundamental frequency
$f_b$. In the gravitational-decay phase of evolution, when energy loss
is dominated by GW emission, the orbit tends to circularise
\cite{an05}. However, at earlier phases of the orbital decay when
three-body stellar interactions or interaction with a gaseous disk
surrounding the binary system are important, the eccentricity may
grow. Stellar three-body interactions can result in orbital decay
through dynamical friction but probably result in a modest increase in
the eccentricity of the black-hole binary system (e.g.,
Ref.~\refcite{mm05}). However, Roedig et al.\cite{rds+11} find that
initially mildly eccentric binary systems decaying through interaction
with a gaseous disk evolve toward a limiting eccentricity in the 0.6
-- 0.8 range. In a regime of frequent mergers it is even possible that
a black-hole triplet could form \cite{ash+10} and, in this case, 
eccentricities as high as 0.99 could exist.

\begin{figure}[ht]
\centerline{\includegraphics[width=85mm]{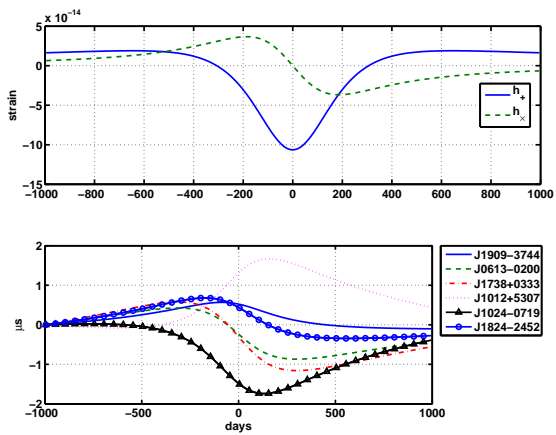}}
\caption{The upper plot shows the gravitational waveforms in the $h_+$
  and $h_-$ polarisations resulting from a parabolic encounter with an
  impact parameter of 0.02 pc of two $10^9 M_\odot$ black holes
  located at a distance of 15 Mpc in the direction of the Virgo
  Cluster. The lower plot shows the resulting timing residuals for
  several PTA pulsars. (Ref.~\protect\refcite{fl10})}
\label{fg:gw_burst}
\end{figure}

Finn and Lommen\cite{fl10} have investigated the GW emission and the resulting
timing residuals from a close parabolic encounter of two massive black
holes. As Figure~\ref{fg:gw_burst} shows, an encounter of two $10^9
M_\odot$ black holes located at a distance of 15 Mpc with a minimum
separation of 0.02 pc produces a GW burst of duration about 1 year and
maximum GW strain $\sim 10^{-13}$. This results in potentially
detectable PTA timing residuals of about $1 \mu$s amplitude with the
same timescale. Unfortunately, the probability of having such a close
encounter of two SMBHs in the local universe within PTA observational
data spans is not high.

Cosmic strings are another potential source of GW burst emission. They
can radiate over a wide frequency range with a huge range of possible
amplitudes and timescales (cf., Figure~\ref{fg:string_spec}) depending
on the detailed mechanism invoked (loops, cusps, short strings, etc.)
and the very wide (virtually unlimited) parameter space
\cite{dv01,smc07,lss09}. Short bursts radiating in the {\it LIGO} and
{\it eLISA} bands may be frequent and unresolved, producing a GWB at
these frequencies. However, bursts with longer timescales, producing
radiation in the nanohertz band, are also possible but are likely to
be extremely rare \cite{dv01}. Consequently, while in principle such
bursts could be detected, in practice it is unlikely that PTAs will be
able to significantly constrain models for GW burst emission from
cosmic strings. 

\subsection{Pulsar Timing Arrays and Current Results}\label{sec:pta}
In this section we first describe the three main PTAs currently
operating world-wide: the European Pulsar Timing Array ({\it EPTA}),
the North American pulsar timing array ({\it NANOGrav}) and the Parkes
Pulsar Timing Array ({\it PPTA}), and the collaboration between them, the
International Pulsar Timing Array ({\it IPTA}). PTAs have many possible
applications such as establishing a pulsar-based timescale
\cite{hcm+12}, investigating the accuracy of solar-system ephemerides
\cite{chm+10}, and investigating the properties of the pulsars
themselves (e.g., Refs~\refcite{ymv+11,sod+14}) and of the intervening
interstellar medium (e.g., Ref.~\refcite{kcs+13}). However, here we
concentrate on what is undoubtedly their primary scientific goal, the
direct detection of gravitational waves. Unfortunately, in common with
other GW detection efforts around the world, PTAs have so far only
been able to place limits on the strength of signals from potential GW
sources. However, these limits are now beginning to seriously
constrain the astrophysical source models and the assumptions that go
into them and hence have implications that go far beyond the GW
studies themselves. 

\subsubsection{Existing PTAs}
The {\it EPTA} uses five large radio-telescopes in Europe, the Effelsberg
100-m telescope in Germany, the Nan\c{c}ay Radio Telescope in France
(95-m equivalent area), the Westerbork Synthesis Radio Telescope in
the Netherlands (similar effective area to the Nan\c{c}ay telescope),
the 76-m Lovell Telescope at Jodrell Bank in England and the recently
completed 64-m Sardinia Radio Telescope in Italy, to observe about 40
MSPs with a cadence of between a few days and 30 days for different
pulsars \cite{kc13}. Different telescopes observe at different
frequencies in the range 0.3 -- 2.6 GHz, but all are instrumented at
1.4 GHz. Normally the five telescopes observe independently, but in a
project known as the ``Large European Array for Pulsars'' ({\it LEAP}) 1.4 GHz
signals over a bandwidth of 128 MHz from the five telescopes can be
summed coherently to form a 194-m equivalent diameter radio
telescope. The different telescopes use different signal-processing
systems, either digital filterbanks or coherent dedisperion systems or
both. 

{\it NANOGrav} makes use of the 300-m Arecibo radio telescope in Puerto Rico
and the 100-m Green Bank Telescope ({\it GBT}) in West Virginia
\cite{mcl13}. A sample of about 36 pulsars is observed,
typically at 3-week intervals. At Arecibo, observations are made in
bands centred at 430 MHz and 1410 MHz, whereas at the {\it GBT}, the
observed bands are centred at 820 MHz and 1500 MHz. Currently both
radio telescopes use coherent dedispersion systems with bandwidths up
to 800 MHz, but in the past a range of filterbank and coherent
dedispersion systems with more limited bandwidths have been used.

As the name suggests, the {\it PPTA} uses the Parkes 64-m radio telescope
located in New South Wales, Australia. A sample of 22 pulsars is
currently being observed with regular observations at 2 -- 3 week
intervals in three bands around 730 MHz, 1400 MHz and 3100 MHz
respectively \cite{mhb+13,hob13}. Coherent dedispersion systems are
used at 730 MHz and 1400 MHz with bandwidths up to 310 MHz and digital
filterbanks at 1400 MHz and 3100 MHz with bandwidths of 256 MHz and
1024 MHz respectively.

\begin{figure}[ht]
\centerline{\includegraphics[angle=270,width=100mm]{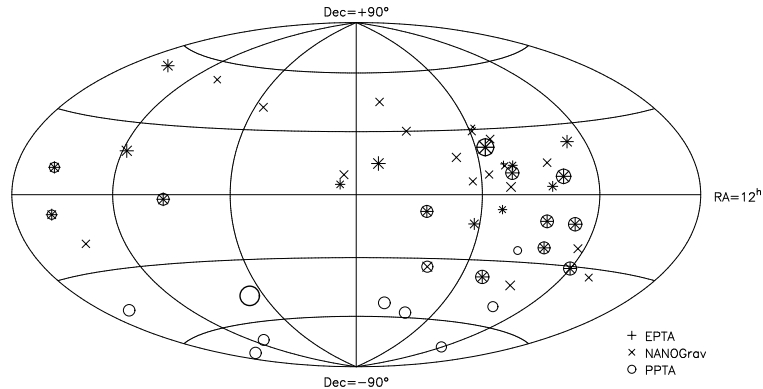}}
\caption{Distribution on the sky of MSPs being timed by the three
  PTAs, with different symbols for each PTA. Right ascension increases
  to the left with $0^{\rm h}$ at the plot centre. The symbol size is
  related to the ratio $S_{1400}/P$, where $S_{1400}$ and $P$ are the
  pulsar 1400 MHz flux density and pulse period
  respectively. (Ref.~\protect\refcite{man13})}
\label{fg:ipta_sky}
\end{figure}

Data sets from all three PTAs have spans ranging from a few years up
to about 20 years for different pulsars; three (including the
original MSP, PSR B1937+21) have Arecibo data spans of nearly 30
years. The three PTAs together observe about 50 pulsars with some
being observed by two or even all three of the PTAs. Their
distribution on the sky is shown in Figure~\ref{fg:ipta_sky}.

Given that the combined data set of the three PTAs contains a larger
number of pulsars, improved observation cadence and greater frequency
diversity than the data set of any one PTA, there is a strong
motivation to combine all the available data sets to obtain maximum
sensitivity for PTA scientific objectives. The International Pulsar
Timing Array ({\it IPTA}) consortium was set up to facilitate progress
toward this goal \cite{man13}. The {\it IPTA} also arranges annual science
meetings and student workshops and provides a forum for outreach
programs and other activities related to PTA research.

\subsubsection{Limits on the nanohertz GW background}
As discussed in Section~\ref{sec:gwb}, GWs from a
cosmological distribution of SMBH binary systems
are expected to contribute a very ``red'' signal to the spectrum of
pulsar timing residuals. The expected signal from other GWB sources is
similar. Consequently, long-term observations of a single pulsar with
little or no detectable intrinsic timing irregularities can be used to
place a limit on the strength of the GWB in the Galaxy. Of course,
statistical limits can be improved by using data from several such
pulsars. An early limit on the GWB at a frequency of 4.5~nHz was set
by Kaspi et al.\cite{ktr94} using Arecibo observations of two MSPs, PSR B1855+09
and PSR B1937+21, with a 95\% confidence limit on $\Omega_{GW}h^2$ of
$6\times 10^{-8}$, where $h=H_0/100$~km~s$^{-1}$.

Since the advent of the various PTA projects, both the quality and
quantity of timing data sets has improved and a variety of analysis
techniques have been employed to extract increasingly restrictive
limits. Based on early {\it PPTA} data on seven pulsars combined with the
Kaspi et al. PSR B1855+09 data set, Jenet et al.\cite{jhv+06} used a
``frequentist'' approach with a statistic based on the amplitude of
the low-frequency components in the power spectrum of the timing
residuals to set a 95\% confidence limit of about $2\times 10^{-8}$ on
$\Omega_{GW}$ at a GW frequency of 1/8~yr or 4~nHz. From Equations
\ref{eq:omgw} and \ref{eq:gwb_spec}, this result is equivalent to a
characteristic strain at frequency 1/1~yr, $A_{1{\rm yr}} \approx
1.1\times 10^{-14}$. 

van Haasteren et al.\cite{vlj+11} analysed {\it EPTA} 1400-MHz data
sets for five MSPs with spans of 5 -- 8 years using a Bayesian
analysis to place limits on the GWB amplitude as a function of its
spectral index $\alpha$. For $\alpha = -2/3$, the derived limit at the
95\% confidence level is $A_{1{\rm yr}} \approx 6\times 10^{-15}$,
about a factor 1.8 better than the Jenet et al.\cite{jhv+06} limit.

{\it NANOGrav} multi-band data sets recorded between 2005 and 2010 for 17
MSPs were analysed by Demorest et al.\cite{dfg+13}. Timing analyses taking into
account time-varying dispersion delays and frequency-dependent pulse
profiles were carried out to form sets of post-fit residuals and the
corresponding covariance matrices for each pulsar. For most MSPs in
the sample, no red noise signal was detectable in the post-fit
residuals. Considering just the pulsar with the smallest post-fit
residuals, PSR J1713+0747, and taking into account absorption of red
noise by the timing fit, Demorest et al. obtained a 95\% confidence
limit for $A_{1{\rm yr}}$ of $1.1\times 10^{-14}$. A separate
cross-correlation analysis weighted by the expected Hellings \& Downs
function (Equation~\ref{eq:hd}) across all the pulsars in the sample
resulted in a somewhat better limit $A_{1{\rm yr}} \approx 7\times
10^{-15}$, although this limit was dominated by correlations with the
two best pulsars in the sample, PSRs J1713+0747 and J1909$-$3744.

\begin{figure}[ht]
\centerline{\includegraphics[width=85mm]{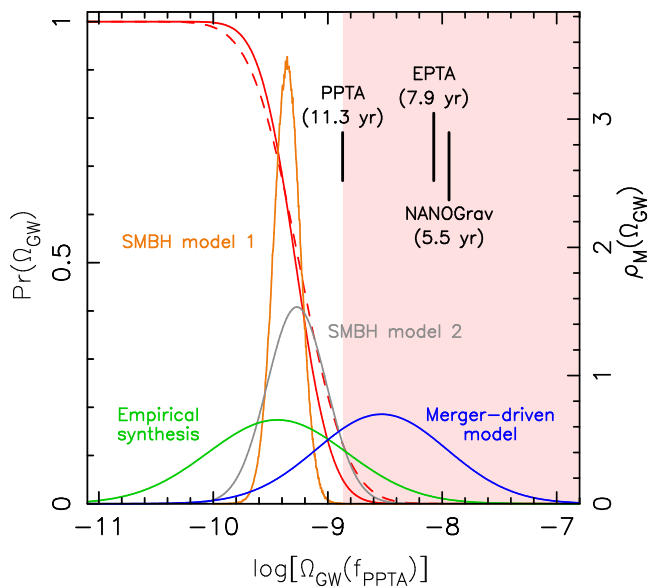}}
\caption{Limits on the relative energy density of the GWB,
  $\Omega_{GW}$ at a GW frequency of 2.8 nHz based on the {\it PPTA}
  data sets, together with predictions for $\Omega_{GW}$ based on
  several different models for the GWB \cite{src+13}. The solid and
  dashed lines that are asymptotic to 1.0 at low $\Omega_{GW}$ show
  the probability $Pr$ that a GWB signal of energy density
  $\Omega_{GW}$ can exist in the {\it PPTA} data sets, based on
  gaussian and non-gaussian GWB statistics respectively. The shaded
  region is ruled out with 95\% confidence by the {\it PPTA}
  data. Corresponding limits from analysis of {\it EPTA}
  \cite{vlj+11} and {\it NANOGrav} \cite{dfg+13} data sets, scaled
  to $f_{GW}$ = 2.8 nHz, are also shown. The gaussian curves show the
  probabilility density functions $\rho_M$ for the existence of a GWB
  with energy density $\Omega_{GW}$ based on a merger-driven model for
  growth of SMBHs in galaxies \cite{mop14}, an empirical synthesis of
  observational constraints on SMBHs in galaxies \cite{ses13}, and
  based on the Millenium dark matter simulations \cite{bsw+09}
  together with semi-analytic models for growth of SMBHs in
  galaxies. (See Ref.~\protect\refcite{src+13} for more detail.)}
\label{fg:ppta_gw}
\end{figure}

Based on {\it PPTA} and earlier Parkes timing observations made in
three observing bands centred near 700 MHz, 1400 MHz and 3100 MHz
respectively with data spans of up to 17 years \cite{mhb+13}, together
with the PSR B1855+09 archival Arecibo data \cite{ktr94}, Shannon et
al.\cite{src+13} placed a limit of $1.3\times 10^{-9}$ on
$\Omega_{GW}$ at a GW frequency of 2.8 nHz. The corresponding limit on
$A_{1{\rm yr}}$ assuming a GW spectral index of $-2/3$ is $2.4\times
10^{-15}$.  This analysis, which included dispersion correction for
the {\it PPTA} data sets and was based on the six best {\it PPTA}
pulsars, used a statistical method similar to that of Jenet et
al.\cite{jhv+06} but included modelling of the red noise in the timing
residuals. As shown in Figure~\ref{fg:ppta_gw}, this limit rules out a
model in which the growth of SMBH in galaxies is dominated by mergers
\cite{mop14} at the 91\% confidence level, but is consistent with
other models for galaxy and SMBH evolution where much of the SMBH
growth is by accretion.

\begin{figure}[ht]
\centerline{\includegraphics[width=85mm]{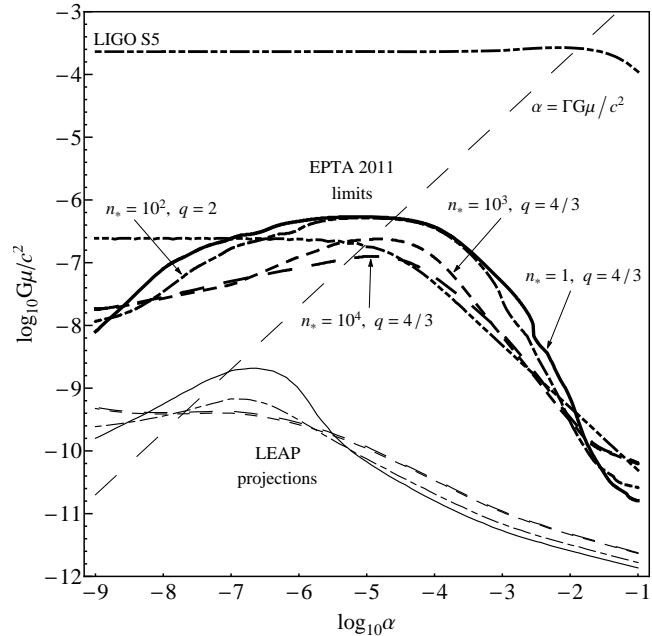}}
\caption{Limits on cosmic string tension as a function of the loop
  size scale parameter $\alpha$ for different cutoff node numbers
  $n_*$ and intrinsic spectral slopes $q$ for the current {\it EPTA}
  limit on the energy density of the GWB (thick lines) and the
  projected sensitivity of the {\it LEAP} PTA (corresponding thin
  lines). The line with one long dash and three short dashes is an
  analytic approximation which is valid for large loops. The uppermost
  line is the {\it LIGO} limit at $f=1$~kHz. 
  (Refs~\protect\refcite{sbs12,aaa+09h})}
\label{fg:string_limits}
\end{figure}

As discussed in Section~\ref{sec:strings}, topological defects in the
early Universe are another potential source for the
GWB. Figure~\ref{fg:string_limits} shows limits on the dimensionless
string tension $G\mu/c^2$ as a function of loop size for various sets
of other relevant parameters \cite{sbs12}. The middle solid curves
are limits based on the current {\it EPTA} data sets and the lower dashed
curves are projections for LEAP data sets that coherently combine data
from the {\it EPTA} telescopes. The upper dot-dashed line is a limit based
on LIGO data \cite{aaa+09h}. The current {\it EPTA} results give a
conservative upper limit on the string tension of $5.3\times
10^{-7}$. Lower limits can be obtained with more restricted
assumptions about the string parameters
(e.g., Refs~\refcite{dv05,jhv+06,vlj+11}).

\subsubsection{Limits on GW emission from individual black-hole binary
  systems}
For an isolated binary system at a luminosity distance $d_L$, the
intrinsic GW strain amplitude is given by
\begin{equation}
h_0 = \frac{(GM_c)^{5/3}}{c^4}\frac{(\pi f)^{2/3}}{d_L}
\end{equation}
where $M_c$ is the binary chirp mass and $f = 2 f_b$ is the GW
frequency. The actual observed signal depends on the orbital
orientation and phase as well as the GW polarisation angle. By
averaging over these quantities, PTAs can set probabilistic limits for
the strain amplitude as a function of $f$, both in a given direction
and averaged over all directions (see, e.g., Refs~\refcite{abb+14,zhw+14}). In
these analyses, there is assumed to be negligible evolution of $f$
over the data span and only the Earth term (cf. Equation~\ref{eq:gwz})
is considered; because of uncertainties in the pulsar distances, the
pulsar terms cannot be added coherently and just contribute noise.

\begin{figure}[ht]
\centerline{\includegraphics[width=85mm]{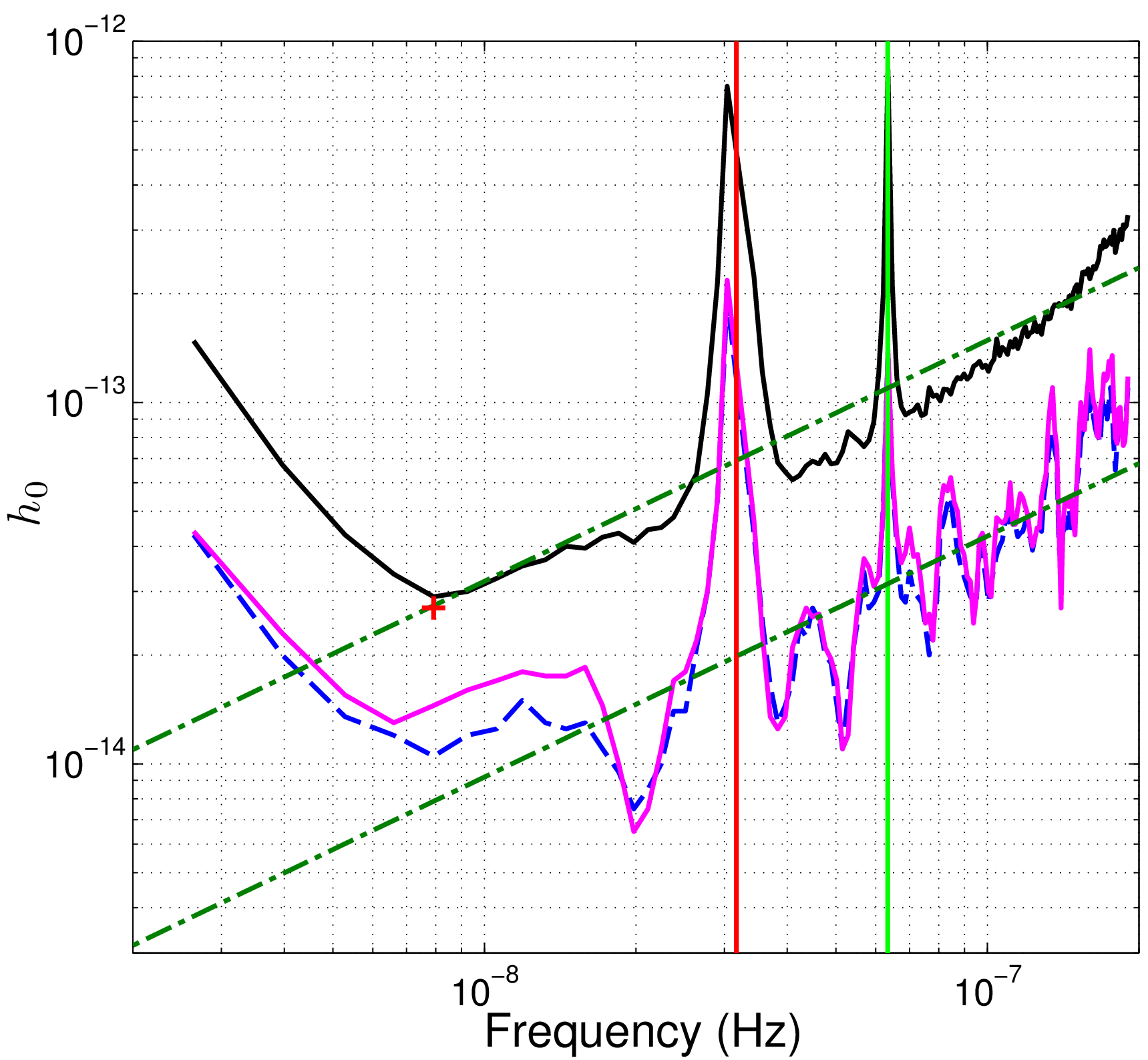}}
\caption{Sky-averaged limits on the intrinsic GW strain amplitude $h_0$ as a
  function of GW frequency $f$ based on the {\it PPTA} data sets. The lower
  curves represent the largest GW signal (with a false-alarm
  probability of 1\%) that could be present in the
  real {\it PPTA} data (dashed line) and a simulated data set (solid
  line). The upper solid line gives the sensitivity of the {\it PPTA} to a
  continuous-wave source, i.e., the minimum signal that could be
  detected with 95\% probability. The upper limits and sensitivities are
  higher at frequencies of 1/1~yr and 1/6~months as these frequencies
  are absorbed by the timing fits for position and parallax
  respectively. The sloping dot-dashed lines are the expected signal levels
  for a SMBH binary systems with $M_c = 10^{10}$~M$_\odot$ and
  distance 400 Mpc (upper line) and $M_c = 10^{9}$~M$_\odot$ and
  distance 30 Mpc (lower line). (Ref.~\protect\refcite{zhw+14})}
\label{fg:gwcw_spec}
\end{figure}

\begin{figure}[ht]
\centerline{\includegraphics[width=120mm]{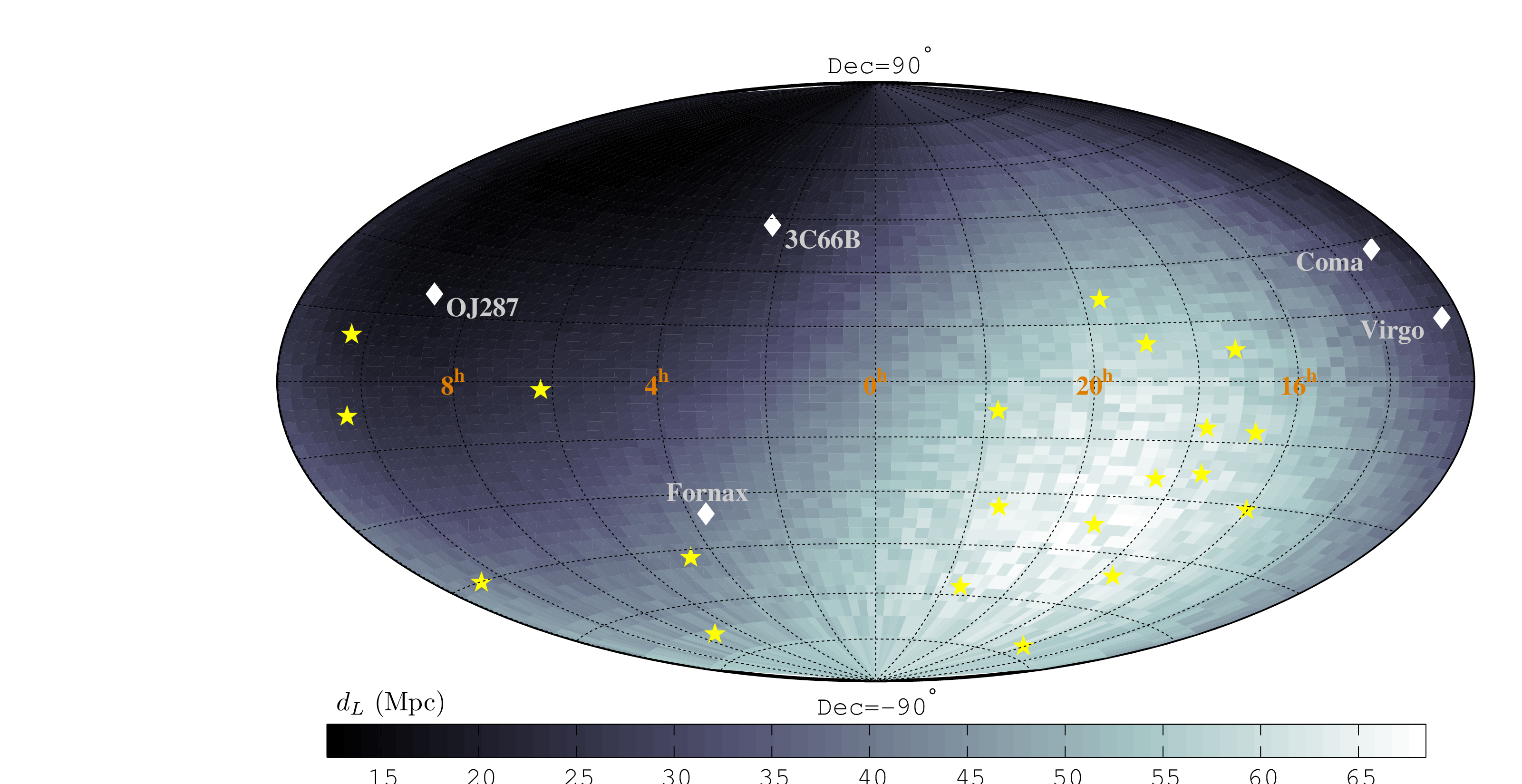}}
\caption{Sky distribution of the luminosity distance $d_L$ to which a
  binary system with chirp mass $M_c = 10^9$~M$_\odot$ radiating at
  $10^{-8}$~Hz could be detected. The stars indicate the positions of
  the 20 {\it PPTA} pulsars and the diamonds are potential sources of GW
  continuous-wave emission. (Ref.~\protect\refcite{zhw+14})}
\label{fg:gwcw_sky}
\end{figure}

Figure~\ref{fg:gwcw_spec} shows both sky-averaged upper limits and
detection sensitivity for continuous-wave GW signals as a function of
GW frequency based on the {\it PPTA} data set and using a frequentist
analysis method \cite{zhw+14}. The best limits and sensitivity are
obtained for GW frequencies around $10^{-8}$~Hz, where the upper limit
on $h_0$ is about $1.1\times 10^{-14}$. A similar analysis of the
five-year {\it NANOGrav} data set by Arzoumanian et al.\cite{abb+14} using
both frequentist and Bayesian analysis methods gave a somewhat higher
sky-averaged upper limit of about $5\times 10^{-14}$ at
$10^{-8}$~Hz. Because of the uneven sky distribution of PTA
pulsars, there is quite a strong dependence of sensitivity on source
direction. This is illustrated in Figure~\ref{fg:gwcw_sky} which shows
that sensitivity is greater toward the greatest concentration of PTA
pulsars, roughly toward the Galactic Centre.

Figure~\ref{fg:gwcw_spec} also shows that we can effectively rule out
the existence of SMBH binary systems with $M_c = 10^9$~M$_\odot$ and
orbital frequencies around $10^{-8}$~Hz at distances closer than
30~Mpc. Similarly, a system with $M_c = 10^{10}$~M$_\odot$ at a
distance of 400~Mpc should be detectable. Unfortunately, as
Figure~\ref{fg:gwcw_sky} shows, the nearby galaxy clusters such as
Virgo, Coma and Fornax are all in regions of relatively low
sensitivity for the {\it PPTA} (and other PTAs), so the effective
limits for these clusters are a factor of a few higher. It is unlikely
that such massive binary systems exist in these clusters. More
generally, limits can also be placed on the SMBH binary coalescence
rate in the nearby Universe ($z\lapp 0.1$). Based on the {\it PPTA}
results, Zhu et al.\cite{zhw+14} place a 95\% confidence limit of
$4\times 10^{-3} (10^{10}{\rm
  M}_\odot/M_c)^{10/3}$~Mpc$^{-3}$~Gyr$^{-1}$ on the coalescence
rate. This limit is about two orders of magnitude above current
estimates of the galaxy merger rate in the local Universe
(cf. Ref.~\refcite{abb+14}).

These rate estimates are based on the orientation-averaged and
sky-averaged amplitudes. It is of course possible that a favourably
oriented and located SMBH binary system in the late stages of
coalescence could exist. On the other hand, the estimates are based on
circular binary orbits and, as discussed in Sections~\ref{sec:gwb} and
\ref{sec:gw_trans} above, short-period SMBH binaries may have
significant eccentricity which reduces the GW power at the fundamental
frequency $f = 2f_b$ and hence the detectability of such systems. On
balance, it seems unlikely that GW from an individual coalescing SMBH
binary system will be detected with the current generation of PTAs. 

\subsection{Future Prospects}\label{sec:future}
PTAs have now achieved data spans and ToA precisions that would allow
detection of the GWB predicted by some models for the evolution of
galaxies and the SMBHs at their core (see, e.g.,
Ref.~\refcite{src+13}). Up to now, no detections have been made. While
this is disappointing from the point of view of GW astrophysics, it is
starting to have important implications for galaxy and SMBH evolution
models and to rule out some scenarios. It also implies that PTAs are
close to detecting the GWB if current predictions for its amplitude
are correct.

\begin{figure}[ht]
\centerline{\includegraphics[width=85mm]{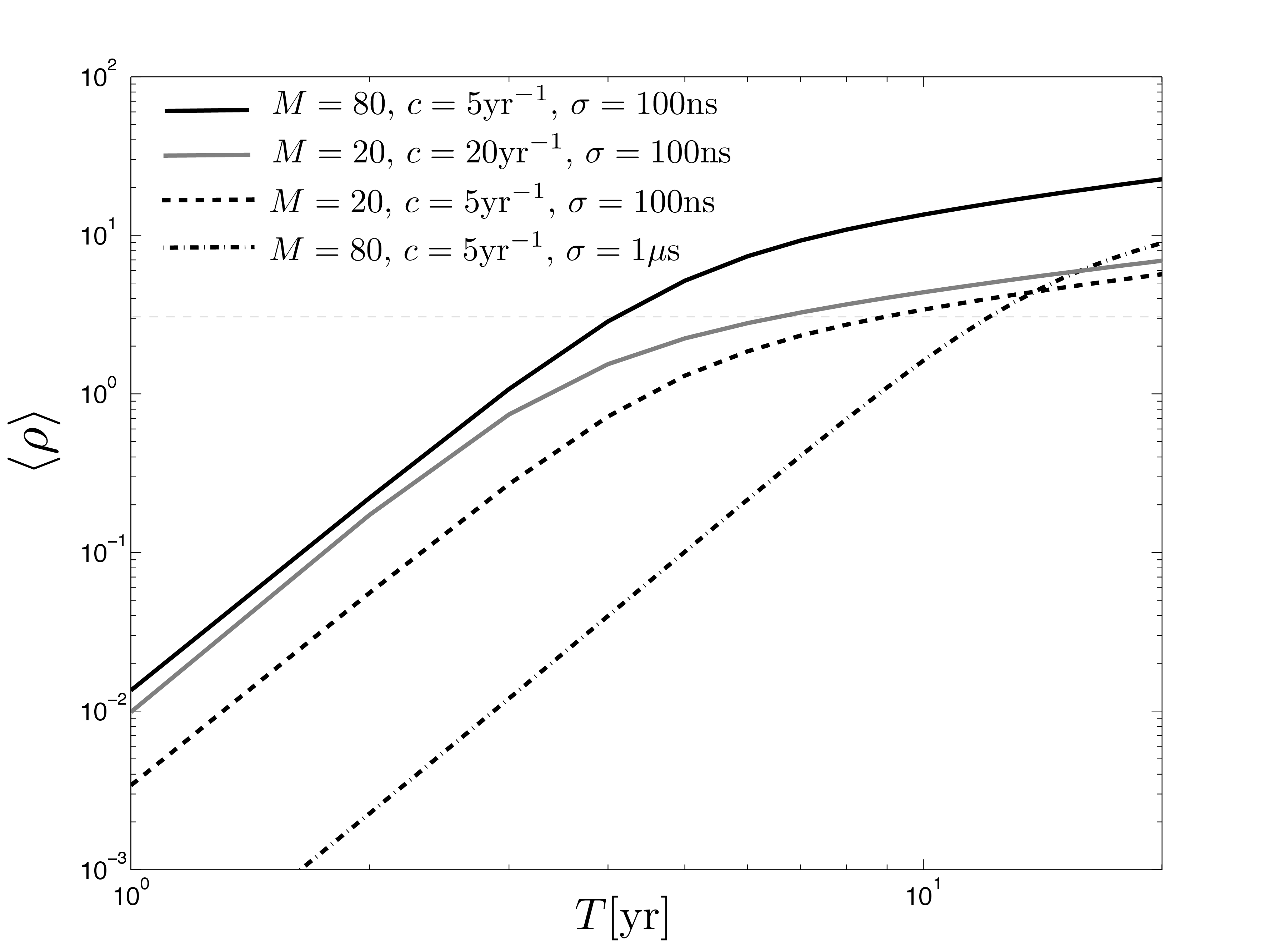}}
\caption{Detection S/N for a GWB as a function of PTA data span for
  four different, but plausible, future PTAs. See text for the meaning
  of the PTA parameters. (Ref.~\protect\refcite{sejr13})}
\label{fg:gwb_sens}
\end{figure}

Siemens et al.\cite{sejr13} have considered the sensitivity of an
idealised PTA to a GWB. At low signal levels, when the lowest signal
frequencies are below the white noise level, the detection
signal-to-noise ratio (S/N) is
\begin{equation}
\langle\rho\rangle \propto Mc\frac{A_{1{\rm yr}}^2}{\sigma^2}T^\beta
\end{equation}
where $M$ is the number of pulsars in the array, $c$ is the observing
cadence (frequency of observations), $A_{1{\rm yr}}$ is the GWB
amplitude (Equation~\ref{eq:gwb_spec}), $\sigma$ is the rms level of
the white timing noise, $T$ is the observing data span and $\beta$ is
the inverse spectral index of the GWB signal in the timing residuals, taken to
be $13/3$ (Equation~\ref{eq:gwb_res}). In the detection regime
where the GWB signal exceeds the white noise level, the S/N is
\begin{equation}
\langle\rho\rangle \propto
M\left(\frac{\sqrt{c}\,A_{1{\rm yr}}}{\sigma}\right)^{1/\beta}T^{1/2}.
\end{equation}
Consequently, in the pre-detection regime, the S/N increases rapidly
with increased observing cadence and data span and decreased timing
noise, but has a much weaker dependence on these parameters in the
strong signal regime. The reason for this is that noise from the
uncorrelated pulsar term, which is also proportional to $A_{1{\rm yr}}$,
dominates over the white ``receiver'' noise, greatly modifying the
statistical behaviour. Importantly though, in both regimes, the S/N is
proportional to $M$, the number of pulsars in the
PTA. Figure~\ref{fg:gwb_sens} illustrates these dependencies for a range
of plausible future PTAs. 

As Siemens et al.\cite{sejr13} point out, if $A_{1{\rm yr}} \sim
10^{-15}$, current PTAs are already in the ``strong signal''
regime. This means that increasing the observing data spans and
cadence or decreasing ToA uncertainties has limited effect on the S/N
of a potential detection. Increasing the number of pulsars in the PTA
is a much more cost-effective way to increase detection
sensitivity. This fact provides much of the motivation to combine data
from existing PTAs to form the International Pulsar Timing Array ({\it
  IPTA}). In the future, the Chinese {\it FAST} radio telescope
\cite{nlj+11} and the {\it SKA} \cite{cr04b} will provide a large
increase in radiometer sensitivity compared to existing
instruments. The discussion above shows that this increased
sensitivity will be best employed in increasing the number of (weaker)
pulsars that are timed, rather than improving the ToA precision on the
stronger existing PTA pulsars. Considerations of ``jitter noise'' in
ToAs resulting from shape variations in individual pulses
\cite{sod+14} lead to the same conclusion.

While direct detection of the GWB would be enormously exciting and
significant, there is no doubt that direct detection of GW from
individual SMBH binary systems is potentially much more interesting
from an astrophysical perspective. It opens up the possibility of
identifying a GW source with source or region identified through
electromagnetic-wave (radio, optical, X-ray or $\gamma$-ray)
emission and the advent of ``multi-messenger'' astronomy
(e.g., Refs~\refcite{lwk+11,srrd12,bur13}). For example, many active galactic
nuclei (AGNs) show evidence of a close binary SMBH at their
core. Examples of this include X-shaped radio lobes
(e.g., Ref.~\refcite{zdw07}), double-peaked or variable emission lines
from the core region (e.g., Ref.~\refcite{sllt13}), quasi-periodic
modulation of core radio (e.g., Ref.~\refcite{vlts11}) or X-ray
emission (e.g., Ref.~\refcite{srrd12,llk14}), and direct imaging of double AGN
(e.g., Ref.~\refcite{rtz+06}). 

With one exception, existing PTA systems do not have sufficient
sensitivity to detect these potential GW sources. The exception is the
claimed SMBH binary system identified by VLBI astrometry of the nearby
quasar 3C~66B \cite{simt03} which was effectively ruled out by pulsar
timing observations \cite{jllw04}. Future PTAs including {\it FAST}
and the {\it SKA} will have much increased sensitivity, making searches
for other GW candidate sources potentially more productive.

Similarly, although blind searches for continuous-wave GW signals in
current PTA data sets have a low probability of successful detection,
in future this should not be the case. The possibility of
identification of a source galaxy or AGN then depends critically on
the accuracy of the position determination for the GW source. When
only detection of the ``Earth term'' is considered, the accuracy is at
best many tens of square degrees (e.g., Ref.~\refcite{zhw+14}) containing
thousands if not millions of galaxies. Only a correlation of the GW
signal with a modulation of some property of a galaxy core (e.g.,
intensity or velocity) would establish an identification.

The situation changes dramatically if the pulsar terms can be added
coherently with the Earth term (cf. Equation~\ref{eq:gwz}). Each
Earth -- pulsar system then forms an interferometer with baseline $d$ and
fringe spacing $\sim \lambda_{GW}/[2\pi d(1+\gamma)]$, where $d$ is
the pulsar distance, $\lambda_{GW}$ is the GW wavelength and $\gamma$
is the direction cosine between the pulsar direction and the GW
propagation direction \cite{lwk+11,bp12}. Positional accuracies are
roughly the fringe spacing divided by the S/N of the GW
detection. Since $d$ is typically $>10^3$ light-years and
$\lambda_{GW}$ is a few light-years, sub-arc-minute positional
accuracies are possible for the stronger sources. However, to achieve
the coherent summation, the distance to the pulsars must be known to
better than $\lambda_{GW}/(1+\gamma)$, i.e., about 1~pc unless the GW
source is nearly aligned with the pulsar. Currently, only one PTA
pulsar, PSR J0437$-$4715, has a distance known to this accuracy,
measured through VLBI astrometry \cite{dvtb08}, but in the {\it SKA}
era this will change. As Boyle and Pen\cite{bp12} point out, with a
high density of PTA pulsars on the sky, advantage can be taken of the
$(1+\gamma)$ factor and so pulsars with less precisely determined
distances located in the general direction of the GW source could be
useful. In this situation, PTAs would not be confusion-limited and, in
principle, many individual GW sources could be identified. With a high
enough number of PTA pulsars (say 1000 or more) it may be possible to
localise a binary GW source by the quadrupolar pattern of timing
residuals in pulsars surrounding the source, even if the pulsar
distances are poorly determined.

\section{Summary and Conclusions}
Nature has been very kind in providing us with a set of near-perfect
celestial clocks, many in situations of rapidly varying gravitational
accelerations. Not only are these celestial clocks, known as pulsars,
precise time-keepers, they are also exceedingly compact. This enables
them to be treated as point masses in theoretical analyses of their
motion and also permits tests in the regime of strong gravitational
fields. These qualities result in a very wide range of applications
for pulsar time-keeping, most importantly, at least in the context of
this review, to investigations of relativistic gravitation.

Observations of double-neutron-star systems, wide circular
pulsar -- white-dwarf systems and even isolated pulsars have been used
to test the accuracy of gravitational theories. Remarkably, general
relativity (GR) is unscathed by all of these tests and hence remains
the most viable theory of gravitation. Pulsar timing has provided the
strongest available limits on at least six parameters describing
deviations from GR. Continued and improved pulsar timing measurements,
especially with new and highly sensitive radio telescopes such as
{\it FAST} and the {\it SKA}, will both improve on these limits and enable new and
different tests of relativistic gravity. They may even
demonstrate a failure of GR to adequately account for the
observations, leading to new or modified theories of gravitation. 

Continuing and new searches for previously unknown pulsars, especially
with {\it FAST} and the {\it SKA}, will not only increase the number of pulsars
that can be used in tests of relativistic gravity. They will also turn
up new and exciting classes of object such as the recently discovered
triple system, PSR J0337+1715. Such discoveries enrich the
investigations of relativistic astrophysics that can be undertaken
with pulsars.

One of the outstanding goals of current astronomy and astrophysics is
the direct detection and study of gravitational waves (GWs). Pulsar
timing arrays (PTAs) provide a viable mechanism for detection of GWs
with frequencies in the nanohertz range. They therefore complement
other existing or planned instruments such as the laser-interferometer
systems {\it LIGO} and {\it eLISA} which are sensitive to GWs at
frequencies of around 100~Hz and millihertz respectively. The most
probable sources for GW detection by PTAs are binary super-massive
black holes in the cores of distant galaxies. These produce an
unresolved background of GWs that is potentially detectable, but there
may also be individual binary systems that could be detected by
PTAs. 

There are currently three major PTAs operating, one each in Europe ({\it EPTA}),
North America ({\it NANOGrav}) and Australia ({\it PPTA}). Up to 
now, no GWs have been detected by PTAs (or other GW detection systems)
so the direct detection of GWs remains a goal. However, recent limits on
the GW background are placing significant constraints on existing
models for galaxy mergers over cosmological time and the formation and
evolution of super-massive black holes in the cores of these
galaxies. For example, a model in which black-hole growth is dominated
by mergers is essentially ruled out. 

The sensitivity of PTAs to GWs is a function of several factors
including the precision of the pulse arrival-time measurements, the
data span of the PTA observations and the cadence or frequency of
observations within this data span. However the most important single
factor is the number of pulsars in the PTA. Of course, these pulsars
must meet certain timing-precision and period-stability criteria to
usefully contribute to a PTA. There are two main approaches to
increasing the number of pulsars. First, existing data sets can be
combined to form a single PTA -- this is the goal of the International
Pulsar Timing Array project. Second, searches can be undertaken to
increase the number of known pulsars suitable for PTA projects. With
{\it FAST} and the {\it SKA} it is possible that hundreds of MSPs that are suitable for PTA
projects will be both discovered and subsequently timed to high
precision. This will surely lead to the detection of GWs and to
detailed investigations of both the GWs themselves and the sources
that generate them.

\section*{Acknowledgments}
I thank my many colleagues in the pulsar community for their insights
and hard work that have led to the results discussed in this
review. In particular, I thank Paulo Freire, Matthew Kerr and Norbert
Wex for helpful comments on earlier versions of the manuscript and
Sydney Chamberlin, Sotiris Sanidas and Xingjiang Zhu for producing
revised versions of figures from their papers. I also thank the CSIRO
and the Australian Research Council for supporting my research over
the years and, in particular, CSIRO Astronomy and Space Science for
their continued support.


\end{document}